\documentstyle[12pt]{article} 
\newcommand{\vsp}{\vspace{3mm}}
\newcommand{\st}{{1\over 16} } 
\newcommand{\half}{\frac{1}{2}}
\newcommand{\hf}{\frac{1}{2}}
\newcommand{\ots}{\otimes}
\newcommand{\ops}{\oplus}
\newcommand{\ts}{\times}

\newcommand{\ol}{\overline}

\newcommand{\qed}{\quad \hbox{\rule[-2pt]{3pt}{6pt}} \par \vspace{3mm}} 
\newcommand{\pr}{\par \vspace{3mm} \noindent {\bf [Proof]} \qquad}
\newcommand{\prend}{\hfill \qed \par \vspace{3mm}}

\newcommand{\owari}{\par \vspace{3mm} }

\newcommand{\1}{\bf 1} 
\newcommand{\bw}{\bf w} 
\newcommand{\al}{\alpha} 
\newcommand{\be}{\beta} 
\newcommand{\de}{\delta}

\newcommand{\ga}{\gamma} 
\newcommand{\la}{\lambda} 
 
\newcommand{\Bbb}{\bf} 
 
\newcommand{\C}{\Bbb C} 
\newcommand{\Z}{\Bbb Z} 
\newcommand{\N}{\Bbb N} 
\newcommand{\Q}{\Bbb Q} 
 
\newcommand{\R}{\Bbb R}

\newcommand{\Ind}{{\rm Ind}}

\newtheorem{thm}{Theorem}[section]
\newtheorem{prn}{Proposition}[section]
\newtheorem{dfn}{Definition}
\newtheorem{lmm}{Lemma}[section]
\newtheorem{cry}{Corollary}[section]

\newtheorem{rmk}{Remark}

\begin{document}
\begin{center}
{\Huge Representation theory of Code VOA and construction of VOAs} \\

\vspace{1cm}

{\large Masahiko Miyamoto} \\

\vspace{5mm}
{ Institute of Mathematics }\\
{ Tsukuba University }\\
{ Tsukuba 305, Japan }\\

\vspace{1cm}
{\large Dedicated to Professor Hiroshi Kimura on his 60th birthday}\\

\end{center}

\begin{abstract}
We study the representation theory of code vertex operator algebras $M_D$ 
(VOAs) constructed from an even binary linear code $D$ in \cite{M2} 
and we then construct VOAs $V$ containing 
a set of mutually orthogonal rational conformal vectors with 
central charge half such that the sum of them is the Virasoro 
element of $V$ using the representation theory of $M_D$. 
The most famous example of such VOAs is the 
Moonshine VOA $V^{\natural}$. If a simple VOA $V$ contains such a set 
of conformal vectors, then $V$ has an elementary 
Abelian automorphism 2-group $P$ generated by involutions given in 
\cite{M1}. 
As a $P$-modules, $V$ has a decomposition 
$V=\oplus_{\chi\in Irr(P)} V^{\chi}$ into the direct sum 
of weight spaces $V^{\chi}$ of $P$. It was proved in \cite{DM}  
that $V^{\chi}$ is an irreducible $V^{1_P}$-module.  
Therefore, the classification of 
such a VOA $V^P$ and the fusion rules of its 
irreducible modules will determine the structure of $V$. 
We will show that 
the fixed point space $V^{1_P}$ is isomorphic 
to a code VOA $M_D$ of some binary linear even code $D$ 
and then study and classify all irreducible 
$M_D$-modules and the fusion rules of them. 
Especially, a Hamming code VOA 
$M_{H_8}$ of $[8,4,4]$-Hamming code $H_8$ 
has a nice property that the tensor product of irreducible 
module with ${\Z}/2$-lowest weights and an irreducible module
is always an irreducible module. 
Namely, the fusion rules of such modules have always a single component.  
Especially, it determines a structure of 
VOA containing a tensor product of Hamming code VOAs by its representations. 
As an application, we show a way of construction of VOA 
from a pair of binary codes satisfying some conditions. 
The key point is that vertex operators of all elements are 
automatically determined and the VOA structure on it is uniquely 
determined. 
\end{abstract}

\renewcommand{\baselinestretch}{1.2}\large\normalsize
\section{Introduction}
The notion of vertex operator algebra (VOA) naturally arises from 
FLM's construction of the moonshine module and Borcherds' insight 
\cite{B}, \cite{FLM}. A VOA is essentially 
the chiral algebra of a two-dimensional conformal quantum field theory.

The most interesting example is the 
moonshine module $V^{\natural}=\sum_{i=0}^{\infty}V_i^{\natural}.$ 
The construction for the moonshine module is treated in 
the book \cite{FLM}.
On the other hand, the simplest example is one 
of the minimal series of rational VOA  
$L(\hf,0)$ with central charge $\half$.  This is generated 
by one rational conformal vector with central charge $\hf$. It is 
known as a critical theory of two-dimensional Ising model 
and it has a relation with the magnetic field. 
Here, a rational VOA means that it has only finitely many 
irreducible modules and any module is a direct sum of 
irreducible modules.

For example, $L(\hf,0)$ has only three 
irreducible modules  $L(\hf,0)$, $L(\hf,\hf)$, and
$L(\hf,\st)$, where the first entry is the central charge and 
the second denotes the lowest weights. 
Its fusion rules are : 
$$ \begin{array}{ll}
(1) &L(\hf,0) \mbox{ is identity},  \\
(2) &L(\hf,\hf)\ts L(\hf,\hf)=L(\hf,0), \\
(3) &L(\hf,\hf)\ts L(\hf,\st)=L(\hf,\st), \mbox{  and }\\
(4) &L(\hf,\st)\ts L(\hf,\st)=L(\hf,0)+L(\hf,\hf).
\end{array}  \eqno{(1.1)} $$

\noindent
The remarkable property of the fusion rules for Ising model is that 
they give us two binary modes:  
$$
\ol{h}:\left\{ \begin{array}{llr}
L(\hf,0),L(\hf,\hf) &\to &0 \\
L(\hf,\st) &\to &1
\end{array} \right. \qquad \qquad
\tilde{h}:\left\{ \begin{array}{llr}
L(\hf,0) &\to &0 \\
L(\hf,\hf) &\to &1 
\end{array} \right. , \eqno{(1.2)} $$
which commute with the fusion rules. \\

Throughout this paper, we will treat a vertex operator algebra $V$ 
containing a set $\{e^1,\ldots,e^n\}$ of mutually orthogonal 
rational conformal vectors $e^i$ with central charge $\hf$ such that 
the sum $\sum_{i=1}^ne^i$ of them is the Virasoro element ${\bf w}$ 
of $V$. 
The most important example is the moonshine VOA $V^{\natural}$. It 
contains mutually orthogonal 48 conformal vectors with central 
charge $\half$ \cite{DMZ}. Since each rational conformal vector 
$e^i$ generates a vertex operator 
subalgebra $<e^i>\cong L(\hf,0)$, $V$ contains a vertex operator 
subalgebra \\
$T\cong \otimes_{i=1}^{n} L(\hf,0)$, that is, the tensor product of 
$n$ copies of $L(\hf,0)$. We can see $V$ as a $T$-module.  
It is also proved in \cite{DMZ} that every irreducible module of tensor
product VOA is a tensor product of each irreducible 
modules.  Namely, every irreducible $T$-module can be expressed in the 
form 
$$\otimes_{i=1}^{n}L(\hf,h^i), \eqno{(1.3)} $$
where $h^i$ is one of $0, \hf, \st$. 

\begin{dfn}
We will call the above n-tuple $(h^1,...,h^n)$ of each lowest weights 
"lowest weight row" of $L$.  We will show that $T$ is rational 
in Corollary 3.1.  Let $W$ be a $T$-module. 
Since $\dim W_m$ is finite and the lowest 
weight $\sum_{i=1}^n h^i$ of 
$\otimes_{i=1}^n L(\hf,h^i)$ is less than or equal to $\hf n$, we obtain that 
$W$ is a finite direct sum of such irreducible $T$-modules.
Let ${\rm lwr}(T,W)$ denote the set of all lowest weight rows 
of irreducible $T$-submodules of $W$ with multiplicities. 
For each irreducible $T$-module $W$ 
with lowest weight row $h=(h^1,...,h^n)$, we assign to it 
a word $\tilde{h}=(\tilde{h}^1,...,\tilde{h}^n)$ by 
$\tilde{h}^i=1$ if $h^i=\st$ and $\tilde{h}^i=0$ otherwise.  
This is given by 
the first binary mode in $(1.2)$.  
We will call this word ``a $\st$-word of $W$'' 
and denote it by $\tilde{h}(W)$. 
Even if $W$ is not an irreducible $T$-module, we 
can use the same notation $\tilde{h}(W)$ whenever 
all irreducible $T$-submodules of $W$ have the same word. 
\end{dfn}

It is proved by the author in \cite{M1} that if a vertex operator 
algebra $V$ contains a rational conformal vector $e$ with central charge 
$\half$, that is, if $V$ contains $L(\hf,0)$, then 
we get an automorphism $\tau_e$ of VOA $V$ by defining 
the endomorphism :
$$ \tau_e\ :\ \left\{
\begin{array}{rl}
1  &\mbox{on $L(\hf,0)$-submodule }U\cong L(\hf,0)\mbox{ or }L(\hf,\hf) \\
-1  &\mbox{on $L(\hf,0)$-submodule }U\cong L(\hf,\st).
\end{array} \right. \eqno{(1.4)} $$
This automorphism is given by the first binary mode in $(1.2)$. 
Applying to our case, we have 
an elementary Abelian automorphism 2-group 
$P=<\!\tau_{e_i}\ |\ i=1,...,n\!>$ generated by 
$n$ mutually commutative automorphisms.
The fixed point space $V^P$ becomes a vertex 
operator subalgebra of $V$ 
and it coincides with the subspace generated by 
all irreducible $T$-modules whose lowest weight rows have no 
$\st$ entries in them.  
Such VOAs are constructed by the author in \cite{M2} as code VOAs $M_D$ 
from even linear codes $D$ 
and we will actually show that $V^P$ is isomorphic to $M_D$ for some 
even binary linear code $D$ in Sec.4.3 if $V^P$ is simple.  We note that 
if $V$ is simple, then so is $V^P$ by \cite{DM}. 
For any linear character $\chi$ of $P$, 
it is proved by Theorem 4.4 in \cite{DM} that the weight space  
$$V_{\chi}=\{v\in V\ |\  gv=\chi(g)v \mbox{ for all }g\in P \} \eqno{(1.5)}$$ 
is a nontrivial irreducible $V^P$-module. 
Therefore, the classification of such code VOAs and their modules becomes 
important for studying these VOAs.  This is our motivation of this paper. 

Since every linear character $\chi$ of $P$ is given by a map 
$$\chi:\tau_{e_i} \to (-1)^{\tilde{h}^i}\quad (\tilde{h}^i=0,1), \eqno{(1.6)}$$
it assigns to $\chi$ 
a binary word $h_{\chi}=(\tilde{h}^1,\ldots, \tilde{h}^n)$. 
Therefore, the component 
$V_{\chi}$ is generated by all irreducible 
$T$-submodules $W$ with $\st$-word $\tilde{h}(W)=h_{\chi}$. 
The set of all $\st$-words $h_{\chi}$ of $T$-submodules 
in $V$ is closed under the sum and hence  
it forms a binary linear code $S$ of length $n$.   We will show that 
$S$ is orthogonal to $D$.

One of the main purposes in this paper is to construct a VOA by the 
reverse way. Namely, assume that we are given a suitable pair of binary 
linear codes $D$ and $S$ satisfying the following conditions: \\

\noindent
{\bf Hypotheses II} \\
(1) $D$ and $S$ are both even linear codes. \\
(2) $\langle D, S\rangle=0 $. \\
(3)  For any $\al\in S$, the weight $|\al|$ is a 
multiple of eight. \\
(4)  For any $\al\in S$, $D$ contains a self-dual subcode $E_{\al}$ 
such that $E_{\al}$ is a direct sum 
$E_{\al}=\oplus_{i=1}^kE^i_{\al}$ of $[8,4,4]$-Hamming 
codes $E_{\al}^i$ and $H_{\al}=\{\be\in E_{\al}:\be\subseteq \al\}$ 
is a direct factor of $E_{\al}$. \\
(5)  For any two codewords $\al, \be\in S$, we assume 
$$K_{\al}+E_{\be}=K_{\al+\be}+E_{\be},  \eqno{(1.7)}$$
where $K_{\al}=\{\be\in D:\be\subseteq \al\}$. \\

\vspace{5mm}

Under the hypotheses II $(1)\sim (5)$, we will first construct a code 
VOA $M_D$. 
One of our assumptions is that $D$ contains a lot of $[8,4,4]$-Hamming codes 
so that we can use the representation theory of Hamming code 
VOA $M_{H8}$, which 
has nice properties as we will see in Sec.5 and Sec.6.  
We will next introduce the concept of induced $M_D$-modules from 
$M_C$-modules for $C\subseteq D$.  
By this concept of induced module, 
we can define an $M_D$-module $V^{\al}$ from a suitable 
$M_{E_{\al}}$-module $U_{\al}$ 
for all $\al\in S$ and assume 
that $V^{\al+\be}$ is a component of the tensor product 
$V^{\al}\times V^{\be}$ for any $\al,\be\in S$. 
We choose $U_{\al}$ so that $M_{E_{\al}}\oplus U_{\al}$ has a simple VOA 
structure on it and assume some commutative condition for 
a pair $(V^{\al}, V^{\be})$.   Set $V=\oplus_{\al\in S}V^{\al}$. 
Under these conditions, we will show that 
the vertex operators $Y(v,z)\in {\rm End}(V)[[z,z^{-1}]]$  
for all $v\in \oplus_{\al\in S}V^{\al}$ are uniquely 
determined up to change of basis and $(V, Y)$ has a VOA structure on it.  \\

In Sec.3, we will recall the properties of VOAs. 
In Sec.4, we will recall a tensor product 
construction of 
code VOA by using a vertex operator superalgebra from \cite{M2}. 
In Sec.5, we will study and classify all irreducible $M_D$-modules. 
In Sec.6, we construct a VOA $V_{H8}$ and study 
the fusion rules. 
In Sec.7, we will construct a VOA 
containing a tensor product $\ots V_{H8}$ and show that a VOA 
structure on it is uniquely determined up to change of basis 
by its representations under some conditions.

\section{Notation}
\begin{tabular}{ll}
$\al=(a^i), \be=(b^i), \ga$ &Words. \cr
$\al^i, \beta^i$ &Words in $({\Z}_2)^8$ or $({\Z}_2)^8/H_8$. \cr
$|\al|$ &The weight of binary word $\al=(a^i)$, 
that is $|\{i\ |\ a^i=1\}|$. \cr
$D, \ K$  &Even binary linear codes. \cr
 $e$   &A conformal vector with central charge $\half$. \cr
 $\{ e^i\ |\ i=1,...,n\}$ &A set of mutually 
orthogonal rational conformal vectors \cr
&with central charge $\half$. \cr
$\{ e^i\ |\ i=1,...,8\}$ &A set of eight coordinate 
conformal vectors in $V_{H8}$. \cr
 $E, \ E_{\al}$  & $E=\ops E^i$. \ Direct sums of Hamming subcodes. \cr
$\{d^i\}, \{f^i\}$ &The other two sets of mutually orthogonal 
eight conformal \cr
&vectors in $M_{H_8}=V_{H8}$. \cr
 $h=(h^i)$ &A lowest weight row.  \cr
$\tilde{h}=(\tilde{h}^i)$ &$\tilde{h}^i=\left\{ \begin{array}{lll}
1 &\mbox{ if } &h^i=\st \cr
0 &\mbox{ if } &h^i=0, \hf 
\end{array} \right. $. A $\st$-word of $h$. \cr
$\tilde{h}(W)$ &The $\st$-word of $W$. \cr 
$H_8$ &The $[8,4,4]$-Hamming code. \cr
$H(\hf,\al)$, $H(\st,\be)$ &The irreducible $V_{H8}$-modules, see Def.13 
in Sec.6.2. \cr
$I\pmatrix{W^1\cr W^2\quad W^3}$ &The space of intertwining operators of type 
$\pmatrix{W^1\cr W^2\quad W^3}$. \cr
$\Ind_{M_E}^{M_D}(U)$ &The induced $M_D$-module from an $M_E$-module $U$, 
see Sec.5.2. \cr
$\hat{K}$ &A central extension $\{ \pm e^k\ |\  k\in K\}$ of 
$K$ by $\pm 1$. \cr
$lwr(T,W)$ &The set of all lowest weight rows, see Def.1. \cr
 $L(\hf,0)$ &The rational Virasoro VOA with central 
charge $\half$. \cr
 $L(\hf,h)$ &An irreducible $L(\hf,0)$-module with lowest weight $h$. \cr
 $M$ &A minimal SVOA $M=M_{\bar{0}}\oplus M_{\bar{1}}$, see Sec.4.1. \cr
 $M_{\bar{0}},\ M_{\bar{1}}$ &The even part and the odd part of SVOA $M$. \cr
 $M_{\al}$ &$=(\otimes_{\al=(a^i)}(M^i)_{\ol{a^i}})\ots e^d$. \cr
 $M_S$ &$=\sum_{\al\in S}M_{\al}$. \cr
$\hat{M}_{\al}$ &$=\otimes_{\al=(a^i)}(M^i)_{\ol{a^i}}$. \cr
$\hat{M}_S$ &$=\sum_{\al\in S}\hat{M}_{\al}$. \cr
 $M_D$ &A VOA constructed from an even binary 
linear code $D$. \cr
\end{tabular} 
\newpage
\begin{tabular}{ll}
 $N$ & A sufficiently large integer. \cr
 ${\N}$ &$=\{0,1,2,\cdots\}$. The set of nonnegative integers. \cr
 $\nu_i$  & $\nu_i=(0^{s-1}10^{n-i})=\{ i\}$. \cr
 $P$ &$=<\!\tau_{e_i}\ |\ i=1,...,n\!>$. \cr
 $q $  &A lowest weight vector 
$q={1\over \sqrt{2}}(e^x+e^{-x})\in M_{\bar{1}}$, see (4.9). \cr
$q^i$ &A copy of $q$ in $M^i$, see (4.33).  \cr
$q^{a^i}$ & $\left\{
 \begin{array}{ll}
 ={\bf 1}^i\in M^i_{\bar{0}} &\mbox{if } a^i=0 \cr
 =q^i\in M^i_{\bar{0}} &\mbox{if } a^i=1.
 \end{array} \right. $ \cr
$q^{\al}$ &$\otimes q^{a^i}$. See (4.33). \cr
$Q$ & A $\hat{K}$-module. \cr
${\Q}$ &The field of rational numbers. \cr
 $\{ s^{\al}\ :\ \al\in {\Z}_2^8/H_8\}$ &The 
other sixteen conformal vectors in $V_{H8}$, see (6.3). \cr
$\sigma_{e}$ &An automorphism of VOA given by $e$ 
of type 2, see \cite{M1}. \cr
Support of $D$ &$=\{i| {}^{\exists}(a^i)\in D \mbox{ s.t. } a^i\not=0 \}$ \cr
$T$  &$=\otimes_{i=1}^nL(\hf,0)$, see Sec.1. \cr
$\tau_{e}$ &An automorphism of VOA given by $e$, see 
\cite{M1}. \cr
$\times$ &A fusion rule or a tensor product. \cr
 $V_{H8}$ &$=M_{H_8}$. The VOA constructed by $H_8$, see Sec.6. \cr
${\bw}, {\bw}^i$ &Virasoro elements. \cr
 $Y(v,z)$& $=\sum_{n\in {\Q}}v_nz^{-n-1}$. A vertex operator of $v$. \cr
${\Z}$  &The ring of integers. \cr
${\Z}_2$ &$=\{0,1\}$. The binary field. \cr
$A(x,y)\sim B(x,y)$ &$\Leftrightarrow 
{}^{\exists} N\in {\N} \mbox{ s.t. }(x-y)^N(A(x,y)-B(x,y))=0$ \cr
$(1^m0^n)$ &$=(1\cdots 10\cdots 0)$ \cr
${\bf 1}, {\bf 1}^i$ &The vacuum. \cr
\end{tabular}

\section{Vertex operators}
\subsection{Properties of VOAs}
Throughout this paper we will treat a VOA $V=\oplus_{i=0}^{\infty}V_n$ 
such that $\dim V_0=1$.  
First, let us recall some basic properties of VOAs. \\

\begin{dfn} Let $V$ be a vector space. We will use the following notations:
$$ \begin{array}{l}
R[[z,z^{-1}]]=\{\sum_{n\in {\bf Z}}a_nz^{-n-1} : a_n\in R \}, \cr
R\{z,z^{-1}\}=\{\sum_{n\in {\bf C}}a_nz^{-n-1} : a_n\in R \}. 
\end{array} $$
\noindent
A weak vertex operator of $V$ is a formal power series \\
$v(z)=\sum_{n\in {\bf Z}}v_nz^{-n-1}\in 
({\rm End} V)[[z,z^{-1}]] $
so that for any $u\in V$ there is an integer $N(u)>\!\!>0$ 
such that $v_nu=0$ for $n>N(u)$.
\end{dfn}

\begin{lmm}[Li, Dong]  Let $V=\sum_{n=0}^{\infty} V_n$ be a Fock space and 
let $u(z)=\sum u_nz^{-n-1}$, \\
 $v(z)=\sum v(m)z^{-m-1}$ 
be weak vertex operators of $V$. 
For any integer $n$, we define the $n$-th normal product by 
$$  u(z)_nv(z)=Res_{z_1}\left((z_1-z)^nu(z_1)v(z)-
(-z+z_1)^nv(z)u(z_1)\right). \eqno{(3.1)} $$
Then $u(z)_nv(z)$ is also a weak vertex operator of $V$.
Here, the binomial coefficients, the binomial expansions and 
the residue ${\bf Res}_{z_1}$  
are defined by 
$$\begin{array}{l}
  \pmatrix{n\cr i}={n(n-1)\cdots (n-i+1)\over i(i-1)\cdots 1}, \cr
 (z_1+z)^n=\sum_{i=0}^{\infty}\pmatrix{n\cr i}{z_1}^{n-i}z^i  ,\cr
 {\bf Res}_{z_1}(\sum a_n{z_1}^{-n-1})=a_{0}.  
 \end{array} $$
We note that $(z_1+z)^n\not=(z+z_1)^n$ in our notation 
if $n$ is not a natural number. Originally, $(z_1+z)^n$ and 
$(z+z_1)^n$ were written by 
$l_{z_1,z}(z_1+z)^n$ and $l_{z,z_1}(z_1+z)^n$.  
\end{lmm}

The next Dong's lemma is important, (see \cite{L1}.) \\

\begin{lmm}[Dong]  Let $u(z), v(z), w(z)$ be weak vertex operators 
of $V$. Assume that 
$u(z), v(z)$ and $w(z)$ satisfy the mutual commutativity, then 
for any integer $n$, $u(z)_nv(z)$ satisfy the commutativity with $w(z)$.
Here the commutativity means 
$$ (z_1-z_2)^Nu(z_1)v(z_2)=(z_1-z_2)^Nv(z_2)u(z_1)  \eqno{(3.2)} $$
for a sufficient large integer $N$.  
\end{lmm}

\vsp

It follows from an easy calculation that 
the normal product also preserves the derivation.

\begin{lmm}  Let $u(z), v(z)$ be weak vertex operators of $V$ and 
$L(-1)\in {\rm End}(V)$.  If $[L(-1),u(z)]={d\over dz}u(z)$ and 
$[L(-1),v(z)]={d\over dz}v(z)$, then \\
$[L(-1),u(z)_nv(z)]={d\over dz}(u(z)_nv(z))$.  
\end{lmm}

We will write only the essential parts of the axioms of VOA, 
its modules, and intertwining operators here.  See \cite{FLM} or the 
others for the detail.

\begin{dfn}
A vertex operator algebra is a ${\bf Z}$-graded vector space 
$V=\sum_{n=0}^{\infty} V_n$ with 
finite dimensional homogeneous spaces $V_n$; equipped with a formal 
power series 
$$Y(v,z)=\sum_{n\in Z} v_nz^{-n-1}\in {\rm End}(V)[[z,z^{-1}]]$$ 
called the vertex operator of $v$ for each $v\in V$ satisfying the 
following $(1)\sim (3)$. \\
(1)  There is a specific element ${\bf 1}\in V_0$ called the 
vacuum such that \\
\mbox{\ }(1.1)  $Y({\bf 1},z)=1_V$ and \\
\mbox{\ }(1.2)  $v_{-1}{\bf 1}=v \mbox{ and }  v_n{\bf 1}=0 
\mbox{ for all } n\geq 0$. \\
(2)  There is an specific element ${\bf w}\in V_2$ called the 
Virasoro element such that \\
\mbox{\ }(2.1)  $\{ L(n):={\bf w}_{n+1}\} $ is a 
Virasoro algebra generator, that is, they 
satisfy 
$$ [L(m),L(n)]=(m-n)L(m+n)+\delta_{m+n,0}{m^3-m\over 12}c,  \eqno{(3.3)}$$
where $c\in {\bf C}$ is called the rank (or the central charge) of $V$, \\
\mbox{\ }(2.2)  the derivation: 
$$  [L(-1), Y(v,z)]={d\over dz}Y(v,z)  \eqno{(3.4)} $$
\mbox{\ }(2.3)  $L(0)_{V_n}=n1_{V_n}$. \\
(3)  The commutativity:  
$$(z-w)^NY(v,z)Y(u,w)=(z-w)^NY(u,w)Y(v,z)\eqno{(3.5)}$$
for a sufficiently large integer $N$. \\
\end{dfn}

\begin{rmk}
An important property of vertex operator is the associativity: 
$$  Y(v_nu,z)=Y(v,z)_nY(u,z).  \eqno{(3.6)} $$
Namely, if we are given vertex operators of $u$ and $v$, then 
vertex operators of $v_nu$ are determined by them. 
Another important property is the skew-symmetry:
 $$  Y(u,z)v=e^{zL(-1)}Y(v,-z)u.  \eqno{(3.7)} $$
Namely, if given the actions of $u$ on $v$, then we obtain those 
of $v$ on $u$. 
One more property we will use is a nondegenerated  
invariant bilinear form $(\ ,\ )$ 
satisfying:
 $$  (v_mu,w)=(u,v_{2k-m-2}w)  \eqno{(3.8)} $$
for $u,v,w\in V$ satisfying $L(0)v=kv$ and $L(r)v=0$ for all $r>0$. 
Since we will treat a VOA $V=\oplus_{i=0}V_n$ with $\dim V_0=1$, 
there is a unique invariant bilinear form 
satisfying $({\bf w},{\bf w})={\rm Rank}(V)/2$ for the Virasoro element 
${\bf w}$ of $V$, see {\rm \cite{L1}}.
\end{rmk}

\vsp

\begin{dfn}  A module for $(V,Y,{\bf 1},{\bf w})$ is a 
${\bf Z}$-graded vector space $M=\oplus_{n\geq 0}M_n$ 
with finite dimensional homogeneous spaces $M_n$; equipped with a formal 
power series 
$$Y^M(v,z)=\sum_{n\in {\bf Z}} v^M_nz^{-n-1}\in ({\rm End}(M))[[z,z^{-1}]] $$ 
called the module vertex operator of $v$ for $v\in V$ satisfying: \\
(1)  $Y^M({\bf 1},z)=1_M$. \\
(2)  $Y^M({\bf w},z)=\sum L^M(n)z^{-n-1}$ satisfies: \\
\mbox{\ }(2.1)  the Virasoro algebra relations, \\
\mbox{\ }(2.2)  the derivation:
$$  Y^M(L(-1)v,z)={d\over dz}Y^M(v,z), and    \eqno{(3.9)}$$
\mbox{\ }(2.3)  $L^M(0)_{M_n}=(k_n)1_{M_n} \mbox{ for some } k_n\in {\bf C}$. \\
(3)  The commutativity. \\
(4)  The associativity: 
$$  Y(u_nv,z)=Y^M(u,z)_nY^M(v,z).  \eqno{(3.10)}$$
\end{dfn}

\subsection{Intertwining operator}

\begin{dfn}  Let $(V,Y,{\bf 1}, {\bf w})$ be a VOA 
and let $(W^1,Y^1)$, $(W^2,Y^2)$ and 
$(W^3,Y^3)$ be three $V$-modules.  An intertwining operator 
of type $\pmatrix{W^1\cr W^2\quad W^3}$ is a linear map 
 $$ \begin{array}{lcll}
I(\ast,z): &W^2 &\to &({\rm Hom}(W^3,W^1))\{z\} \cr
 &u &\to &I(u,z)=\sum_{n\in {\Q}} u_nz^{-n-1}
\end{array}  $$
satisfying: \\
(1) Derivation:
$$  I(L^1(-1)u,z)={d\over dz}I(u,z).\eqno{(3.11)}$$
(2) The commutativity:  for $v\in V, u\in W^2$, 
 $$  (z-z_1)^N\{ Y^1(v,z)I(u,z_1)-I(u,z_1)Y^3(v,z) \}=0 \eqno{(3.12)} $$
for a sufficiently large integer $N$. \\
(3) The associativity:
$$ I(v^1_nu,z)=Y(v,z)_nI(u,z), \eqno{(3.13)}$$
where the normal product 
$Y(v,z)_nI(u,z)$ is given by 
$$ {\bf Res}_{z_1}\{(z_1-z)^nY^1(v,z_1)I(u,z)-(-z+z_1)^nI(u,z)Y^3(v,z_1)\} 
\eqno{(3.14)}$$
and $Y^i({\bf w},z)=\sum_{n\in {\bf Z}}L^i(n)z^{-n-2}$.   
\end{dfn}

\vsp

\begin{dfn}
$I_V\pmatrix{W^1\cr W^2\quad W^3}$ denotes the set of 
intertwining operators of type \\
$\pmatrix{W^1\cr W^2\quad W^3}$. It is a 
vector space and its dimension is denoted by $N^{W^1}_{W^2,W^3}$. 
In order to denote the dimensions, we use an expression
 $$  W^2\ts W^3=\sum_{W} N^{W}_{W^2,W^3}W,   \eqno{(3.15)} $$
called ``fusion rule'', where $W$ runs over all irreducible $V$-modules. 
\end{dfn}

\begin{dfn}
To simplify the notation, we sometimes omit $V$ in 
$I_V\pmatrix{W^1 \cr W^2 \quad W^3}$.
\end{dfn}

We recall the definition of a tensor product from \cite{L2}.

\begin{dfn}  Let $M^1$ and $M^2$ be two $V$-modules.  A pair 
$(M, F(\ast,z))$, which consists of a $V$-module $M$ and an 
intertwining operator $F(\ast,z)$ of type 
$\pmatrix{M\cr M^1\quad M^2}$, is called a tensor product for 
the ordered pair $(M^1,M^2)$ if the following universal property holds:
For any $V$-module $W$ and any intertwining operator $I(\cdot,z)$ of 
type $\pmatrix{U\cr M^1\quad M^2}$, there exists a unique 
$V$-homomorphism $\psi$ from $M$ to $U$ such that 
$I(\ast,z)=\psi\cdot F(\ast,z)$. Here $\psi$ extends canonically 
to a linear map from $M\{ z\}$ to $U\{ z\}$ and 
$U\{z\}$ denotes the set of all formal power series 
$ \sum_{n\in {\C}}u_nz^n $ with $u_n\in U$. 
\end{dfn}

\begin{rmk}  We should note that we can't define a tensor product 
of two modules generally. However, since we will treat only rational VOAs 
in this paper, 
the tensor products of two modules $M^1$ and $M^2$ are always well-defined 
and it is isomorphic to 
$\oplus_UN^{U}_{M^1,M^2}U, $
where $U$ runs over the all irreducible $V$-modules.
Therefore, it is equal to the fusion rule in our case and so 
we will use the same notation $M^1\times M^2$ to denote the tensor product.  
It is known that 
$M^1\times M^2\cong M^2\times M^1$. 
\end{rmk}

\subsection{tensor products of VOAs} 
Let $(V^1, Y^1, {\bf 1}^1, {\bf w}^1)$ and 
$(V^2, Y^2, {\bf 1}^2, {\bf w}^2)$ be VOAs.  
For $v^1\otimes v^2\in V^1\otimes V^2$, define the vertex operator 
$$(Y^1\otimes Y^2)(v^1\otimes v^2,z)=Y^1(v^1,z)\otimes Y^2(v^2,z)\in 
{\rm End}(V^1\otimes V^2)[[z,z^{-1}]]$$ 
of $v^1\otimes v^2$ and extend it linearly to 
the whose space $V^1\otimes V^2$, then 
$$(V^1\otimes V^2, Y^1\otimes Y^2, {\bf 1}^1\otimes {\bf 1}^2, {\bf w}^1\otimes {\bf 1}^2
+{\bf 1}^1\otimes {\bf w}^2) \eqno{(3.16)}$$ becomes a VOA.  
We will call it the tensor product of $V^1$ and $V^2$.  

It is known in \cite{DMZ} and \cite{FHL} 
that irreducible $V^1\otimes V^2$-modules 
are tensor products of each irreducible modules.  

We will study the fusion rules of $V^1\otimes V^2$-modules.  
Assume that $V^1$ and $V^2$ are rational, that is, they have only 
finitely many irreducible modules and their modules are direct sums of 
irreducible submodules. 

\begin{thm}  Assume that $V$ and $W$ are rational VOAs.  
Let $V^1$, $V^2$, $V^3$ be irreducible $V$-modules and 
$W^1$, $W^2$, $W^3$ irreducible $W$-modules such that 
$N^{V^3}_{V^1\ V^2}\leq 1$.  Then 
$$
I_V\pmatrix{V^3\cr V^1\quad V^2}\otimes I_W\pmatrix{W^3\cr W^1\quad W^2} 
=I_{V\otimes W}\pmatrix{V^3\otimes W^3 \cr V^1\otimes W^1\quad V^2\otimes W^2}
\eqno{(3.17)}$$
\end{thm}

\pr
We will use the same argument as in the proof of Proposition 4.4 in \cite{M1}. 
Clearly, we have 
$$
I\pmatrix{V^3\cr V^1\quad V^2}\otimes I\pmatrix{W^3\cr W^1\quad W^2} 
\subseteq 
I\pmatrix{V^3\otimes W^3 \cr V^1\otimes W^1\quad V^2\otimes W^2}.  
$$
We will show the reverse.  Take  $$I(\ast,z)\in 
I\pmatrix{V^3\otimes W^3 \cr V^1\otimes W^1\quad V^2\otimes W^2}.$$  
Set $T=V\otimes W$ and view $V$ and $W$ as subVOAs 
$V\otimes {\bf 1}_W$ and 
${\bf 1}_V\otimes W$, respectively. Let $e$ and $f$ be 
Virasoro elements of $V$ and $W$.  Since $V$ is rational, 
$V^i\otimes W^i$ are direct sums of 
irreducible $V$-modules.  
Let $M^1$, $M^2$, $M^3$ be irreducible $V$-submodules of 
$V^1\otimes W^1$, $V^2\otimes W^2$, and $V^3\otimes W^3$, respectively.  
Clearly, $M^i$ is isomorphic to $V^i$ as $V$-modules for $i=1,2,3$. 
Let $\pi$ denote the projection map $\pi: V^3\otimes W^3 \to M^3$ 
such that 
$\pi_{|M^3}=1_{M^3}$.  
Since $f_1$ acts on $W^i$ diagonally and the actions of $f_1$ on 
$V^i\otimes W^i$ 
commutes with the actions of $V$ on $V^i\otimes W^i$, we may assume that 
$f_1$ acts on $M^i$ as scalars $\al_i\in {\bf C}$.   
Let $u\in M^1$, $v\in M^2$. Since 
$$ \begin{array}{l}
 f^3_1u_{m-1}v=\{f^3_1u_{m-1}-u_{m-1}f^2_1\}v+u_{m-1}f^2_1v \cr
 =(f_0u)_mv+(f_1u)_{m-1}v+u_{m-1}\al^2v  \cr
 =(f_0u)_mv+(\al^1+\al^2)u_{m-1}v, 
\end{array} $$
 we get
 $$ (f_0u)_mv=(f_1-\al^1-\al^2)u_{m-1}v $$
 and also 
 $$ (f_0u)_mv=((L(-1)-e_0)u)_{m}v=(L(-1)v)_mu-(e_0v)_mv 
 =-mu_{m-1}v-(e_0u)_mv.  $$
We hence have 
$$ \pi ((f_0u)_mv)=(\al^3-\al^1-\al^2)\pi(u_{m-1}u).  $$
For simplicity, set $\al=\al(M_2,M_3)=\al^3-\al^1-\al^2$.  Then we obtain 
$$ \pi((e_0u)_mv)=\pi(((L(-1)-f_0)u)_mv)=(-m-\al)\pi(u_{m-1}v).  $$
If we set $I^1(u,z)=\pi(I(u,z))z^{-\al}$, then it is easy to see that 
$I^1(\ast,z)$ satisfies the Jacobi identity.  Also, the above equation 
implies the $e_0$-derivative formula: 
$$ I(e_0u,z)v=({d\over dz}I^1(u,z))v.  $$
Hence, $I^1(\ast,z)$ is an intertwining operator of $V$ of type 
$\pmatrix{V^3\cr V^1\quad V^2}$.   
We fixes an intertwining operator $I^1(\ast,z)$ of type 
$\pmatrix{V^3\cr V^1\quad V^2}$. 
By the assumption $N^{r^3}_{r^1\ r^2}\leq 1$, 
we have an expression $I(u,z)=B(w)\otimes I^1(u,z)$ 
for $w\in W^1$ and $u\in M^1=V\otimes w$, where 
$B(w)=(\la_{M^2,M^3}z^{\al(M^2,M^3)})$ is an infinite dimensional matrix, 
where $M^2$ and $M^3$ run over all irreducible components of the 
direct sums of $V^2\otimes W^2$ and 
$V^3\otimes W^3$.    
In particular, we may view $B(w)$ as an element of 
${\rm Hom}(W^2,W^3)\{z,z^{-1}\}$.  
Since $I(\ast,z)$, $I^1(\ast,z)$ satisfy the commutativity, the associativity, 
and the derivation, so does $B(w)$.  Hence 
we conclude that $B(w)=I^2(w,z)$ is an intertwining 
operator of $W$ and  $I(u\otimes w,z)=I^1(u,z)\otimes I^2(w,z)$ 
as desired. 
\prend

As a corollary, we will show the following: 

\begin{cry}  Assume that $V^1$ and $V^2$ are rational VOAs, then 
$V^1\otimes V^2$ is rational. 
\end{cry}

\pr 
Set $V=V^1\otimes V^2$. 
Suppose false and let $(W,Y^W)$ be an indecomposable $V$-module. 
Since $V$ has only finite number of nonisomorphic 
irreducible modules by \cite{DMZ}, 
$W$ has an irreducible $V$-submodule $W^0=U^1\otimes U^2$. 
Since $V^i\times U^i\cong U^i$ for $i=1,2$, we have that 
all irreducible factors of $W$ are isomorphic to $W^0$ and hence 
$W$ has a series $0\subseteq W^0\subseteq \cdots \subseteq W^k=W$ 
such that $W^i/W^{i-1}\cong W^0$ as $V$-modules.  
In particular, $W^i=W^{i-1}\oplus X^i$ splits as a vector space and 
$W=\oplus_{i=0}^k X^i$ as vector spaces. 
Let $\pi$ be a projection $\pi^i:W^i \to X^i$ such that $\pi(W^{i-1})=0$.  
Then for $v^1\in V^1$, $v^2\in V^2$, by taking 
the restriction into $X^i$ and the image of $\pi^j$ of the  
module vertex operator, we have  
$$  \pi^jY(v^1\otimes v^2,z):X^i \to X^j\in 
{\rm Home}(X^i,X^j)U^j[[z,z^{-1}]].  $$
It follows from the properties of the module vertex operator that 
the above is an intertwining operator of type 
$\pmatrix{U^1\otimes U^2 \cr V^1\otimes V^2 \qquad U^1\otimes U^2}$. 
Since $V^1\times W^0=W^0$ and $V^2\times U^0=U^0$, we have 
$\dim I\pmatrix{U^1\otimes U^2 \cr V^1\otimes V^2 \qquad U^1\otimes U^2}=1$ 
by the previous theorem. 
Let  \\
$I(\ast,z)\in I\pmatrix{U^1\otimes U^2 \cr V^1\otimes V^2 
\qquad U^1\otimes U^2}$ be a nonzero intertwining operator such that \\
$I({\bf 1},z)=id_{U^1\otimes U^2}$. Since $W=\otimes_{i=0}^k X^i$, we have 
the following expression:
$$  Y(v,z)=A \otimes I(v,z),  $$
for some $(k+1)\times (k+1)$-matrix $A$.  Since $Y({\bf 1},z)=1_W$, 
$A$ is the identity matrix and $W$ is a direct sum of irreducible 
$V^1\otimes V^2$-modules.
\prend

\section{Tensor product of SVOAs}
In this section, we recall the results from \cite{M2} and 
construct a VOA $M_D$ from an even binary linear code $D$.  

\subsection{Minimal SVOA M}
Let us construct first a vertex operator superalgebra (SVOA)
 $$ M=L(\hf,0)\ops L(\hf,\hf). \eqno{(4.1)}$$
Let $L={\Z}x$ be a lattice with $\langle x,x\rangle=1$. 
Viewing $H={\bf C}x$ as a commutative Lie algebra with a bilinear form 
$<,>$, we define the affine Lie algebra 
$$\hat{H}=H[t,t^{-1}]+{\bf C}k$$ 
associated with $H$ 
and the symmetric tensor algebra $M(1)$ of $\hat{H}$.  As they did in 
\cite{FLM}, we shall define 
the Fock space $V_L=\oplus_{a\in L}M(1)e^a$ with the vacuum ${\bf 1}=e^0$ 
and the vertex operators $Y(\ast,z)$ as follows: 

The vertex operator of $e^a$ is given by 
$$ Y(e^a,z)=exp
\left(\sum_{n\in {\Z}_+}{a(-n)\over n}z^{n}\right)exp
\left(\sum_{n\in {\Z}_+}{a(n)\over -n}z^{-n}\right)
e^{a}z^{a}. \eqno{(4.2)} $$
and that of $a(-1)e^0$ is 
$$ Y(a(-1)e^0,z)=\sum a(n)z^{-n-1}.$$
The vertex operators of other elements are defined by the normal product:\\
$$  Y(a(n)v,z)=Y(a(-1)e^0,z)_nY(v,z).  \eqno{(4.3)}$$

By the direct calculation as they did in \S 4 in \cite{FLM}, we have: 

\begin{lmm}
$$ \begin{array}{ll}
 [\sum a(n)x^{-n-1}, Y(e^b,z)]=0 &\mbox{ for any } 
 a,b\in L,  \cr
(x-z)^N[Y(e^a,x),Y(e^b,z)]=0 &\mbox{ for } <a,b>\equiv 2 \pmod{2},  \cr
(x-z)^N\{Y(e^a,x)Y(e^b,z)+Y(e^b,z)Y(e^a,z)\}=0 &\mbox{ for } <a,b>\equiv 1 
\pmod{2},  
\end{array} \eqno{(4.4)} $$
for a sufficiently large integer $N$.  
These are called "supercommutativity". 
\end{lmm}

\begin{rmk}
We note that the above relations hold generally. Namely, 
if we define a module vertex operator $I^W$ on $W=M(1)\otimes e^{\hf {\Z}x}$ 
by 
$$ I^W(e^a,z)=exp
\left(\sum_{n\in {\Z}_+}{a(-n)\over n}z^{n}\right)exp
\left(\sum_{n\in {\Z}_+}{a(n)\over -n}z^{-n}\right)
e^{a}z^{a} \eqno{(4.5)}$$
and 
$$ I^W(a(-1){\bf 1},z)=\sum a(n)z^{-n-1},  \eqno{(4.6)} $$
then they satisfy the same super-commutativity. 
We should note that if $a\not\in 2{\bf Z}x$, then the powers of $z$ 
in $I^W(e^a,z)$ are not integers since $z^xe^{x/2}=z^{\hf}$. 
Namely, $I^M(e^a,z)$ does not 
satisfy one of the conditions of module vertex operators. 
\end{rmk}

In \cite{M2}, we found two mutually orthogonal conformal vectors 
$$\begin{array}{ll}
{\bw}^1={1\over 4}x(-1)^2{\bf 1}+{1\over 4}(e^{2x}+e^{-2x})  &\mbox{ and }\cr
{\bw}^2={1\over 4}x(-1)^2{\bf 1}-{1\over 4}(e^{2x}+e^{-2x})  & 
\end{array} \eqno{(4.7)}$$
with central charge $\half$ such that 
${\bw}={\bw}^1+{\bw}^2={1\over 2}x(-1)^2{\bf 1}$ is the Virasoro 
element of $V_L$, where ${\bf 1}=e^0$ is the vacuum.
We also note that ${\bw}^1$ and ${\bw}^2$ are rational since 
$V_L$ has a positive definite invariant bilinear form induced from 
the inner product of $L$.
As we mentioned at the beginning,  
$<\!{\bw}^i\!>\cong L(\hf,0)$ has only three irreducible modules 
$L(\hf,0), L(\hf,\hf),L(\hf,\st)$. 
By calculating the dimensions of weight spaces, 
we conclude that $V_L$ is isomorphic to the tensor product 
$ \left( L(\hf,0)\ops L(\hf,\hf)\right)
\ots\left(L(\hf,0)\ops L(\hf,\hf)\right) $
as $<\!{\bw}^1\!> \ots <\!{\bw}^2\!>$-modules, where 
$<\!{\bw}^i\!>$ denotes the vertex operator subalgebra 
generated by ${\bw}^i$. 
Let $\theta$ be the automorphism of $V_L$ induced from the automorphism 
$-1$ on $L$. Take 
the fixed point space $(V_L)^{\theta}$ of $V_L$ by $\theta$. 
This is also a SVOA containing ${\bw}^1$ and ${\bw}^2$ and 
we obtain the decomposition:
$$(V_L)^{\theta}\cong \left(L(\hf,0)\ots L(\hf,0)\right)\quad \ops \quad 
\left(L(\hf,\hf)\ots L(\hf,0)\right) \eqno{(4.8)}$$
as $<\!{\bw}^1\!>\ots <\!{\bw}^2\!>$-modules by 
calculating the dimensions of weight spaces.
Take the subspace $M=\{ v\in (V_L)^{\theta}\ |\  {{\bw}^2}_1v=0\}$. Then 
$M$ is a SVOA with the Virasoro element ${\bw}^1$ and we see 
$M\cong L(\hf,0)\ops L(\hf,\hf) $
as $<\!{\bw}^1\!>$-modules. \\

Therefore, we proved the following theorem (see \cite{M2}).\\

\begin{thm} $(M,Y,{\bf w}^1,e^0)$ is a simple SVOA with the even part
 $M_{\ol{0}}\cong L(\hf,0)$ and the odd part 
$M_{\ol{1}}\cong L(\hf,\hf)$ as $L(\hf,0)$-modules.
The central charge of $(M,Y)$ is $\half$ and it has a positive definite 
invariant bilinear form.  Here we define the vertex operator $Y$ 
by restricting $V_L$ into $M$. 
\end{thm}

For the later use, we fix a lowest weight vector 
$$ q={1\over \sqrt{2}}(e^x+e^{-x})\in M_{\bar{1}} \eqno{(4.9)} $$ 
of the odd part.  By the direct calculation, we obtain 
$$  q_{-2}q=2{\bw}^1, \quad q_{-1}q=0, \quad q_0q={\bf 1}. \eqno{(4.10)} $$

It is easy to check the following correspondences in $V_{{\Z}a}$: 
$$ \begin{array}{lll}
x(-1)e^0 &\in &L(\hf,\hf)\ots L(\hf,\hf),  \cr
{1\over \sqrt{2}}(e^x+e^{-x}) &\in   &L(\hf,\hf)\ots L(\hf,0), \mbox{  and} \cr
{1\over \sqrt{2}}(e^x-e^{-x}) &\in   &L(\hf,0)\ots L(\hf,\hf). 
\end{array} \eqno{(4.11)} $$
We also get ${1\over 2}(e^x+e^{-1})_{-1}(e^x-e^{-1})=x(-1)e^0$. \\

We can construct a $V_{2{\Z}x}$-module 
$W=V_{{\Z}x/2}$ and the module vertex operator $Y^W$ 
by the same ways. As we mentioned in Remark 3, 
we have a formal power series $I^W(v,z)\in {\rm End}(W)[[z^{1/2},z^{-1/2}]]$ 
for $v\in V_{{\Z}x}$. 
Clearly, for $v\in V_{2{\Z}x}$, $I^W(v,z)=Y^W(v,z)$ and hence 
$I^W(q,z)=\sum q_nz^{-n-1}$ satisfies the 
commutativity with $Y^W(u,z)$ of $u\in V_{2{\Z}x}$. 
Namely, $I^W(\ast,z)$ is an intertwining operator of type 
$\pmatrix{W \cr M \quad W}$. 
The $V_{2{\Z}x}$-module $V_{{\Z}x/2}$ splits up into a 
direct sum 
$$V_{{\Z}x/2}=V_{{\Z}x}\ops V_{2{\Z}x+x/2}\ops V_{2{\Z}x-x/2} $$
of irreducible $V_{2{\Z}x}$-modules 
and $q_n$ permutes $\{ V_{2{\Z}x+x/2}, V_{2{\Z}x-x/2} \}$ 
for any $n=\hf+{\Z}$.
By the direct calculations, we see that $V_{2{\Z}x+x/2}$ 
and $V_{2{\Z}x-x/2}$ are both isomorphic to 
$L(\hf,\st)\ots L(\hf,\st)$ as 
$<\!{\bw}^1\!>\ots<\!{\bw}^2\!>$-modules. 
Fix the lowest weight vectors $e^{\hf x}$ and $e^{-\hf x}$ 
of $V_{2{\Z}x+x/2}$ and $V_{2{\Z}x-x/2}$, respectively. 
By restricting $v$ in $M_{\bar{1}}\cong L(\hf,\hf)$ and taking 
the eigenspaces of ${\bf w}^2$ with an eigenvalue $\st$,  
$I^W(v,z)$ defines the following three intertwining operators: 
$$\begin{array}{l}
I^{\hf,0}(\ast,z)\in I\pmatrix{\hf\cr \hf \quad 0},   \cr 
I^{\hf,\hf}(\ast,z)\in I\pmatrix{0 \cr \hf \quad \hf} \mbox{ and }\cr
I^{\hf,\st}(\ast,z)\in I\pmatrix{\st \cr \hf \quad \st}. 
\end{array} \eqno{(4.12)} $$

Also, the restriction to $M_{\bar{0}}\cong L(\hf,0)$ 
defines the following intertwining operators: \\
$$ \begin{array}{ll}
I^{0,0}(\ast,z)\in I\pmatrix{0 \cr 0 \quad 0}, &  \cr 
I^{0,\hf}(\ast,z)\in I\pmatrix{\hf \cr 0 \quad \hf} \mbox{ and }\cr
I^{0,\st}(\ast,z)\in I\pmatrix{\st \cr 0 \quad \st}, 
\end{array}   
\eqno{(4.13)}$$
which are actually module vertex operators of $<{\bf w}^1>$.
We fix these intertwining operators throughout this paper. 

Using the above intertwining operators, the formal power series  \\
$Y(u\otimes v,z)\in {\rm End}(V_{(\hf+{\bf Z})x})\{z,z^{-1}\}$ 
given by (4.2) and (4.5) are 
$$ \begin{array}{ll}
Y(u\otimes v,z)
=\pmatrix{1&0\cr 0&1}\otimes I^{0,\st}(u,z)\otimes I^{0,\st}(v,z) 
&\mbox{  for } u\otimes v\in L(\hf,0)\otimes L(\hf,0), \cr
Y(u\otimes v,z)
=\pmatrix{1&0\cr 0&-1}\otimes I^{\hf,\st}(u,z)\otimes I^{\hf,\st}(v,z) 
&\mbox{  for } u\otimes v\in L(\hf,\hf)\otimes L(\hf,\hf), \cr
Y(u\otimes v,z)
=\pmatrix{0&1\cr 1&0}\otimes I^{\hf,\st}(u,z)\otimes I^{0,\st}(v,z) 
&\mbox{  for } u\otimes v\in L(\hf,\hf)\otimes L(\hf,0), \cr
Y(u\otimes v,z)
=\pmatrix{0&1\cr -1&0}\otimes I^{0,\st}(u,z)\otimes I^{\hf,\st}(v,z) 
&\mbox{  for } u\otimes v\in L(\hf,0)\otimes L(\hf,\hf). 
\end{array} \eqno{(4.14)}$$

\begin{rmk}
By the definitions (4.2) and (4.5) of $Y(\ast,z)$ and $I^W(\ast,z)$, 
we have: \\
(1) The powers of $z$ in $I^{0,\ast}(\ast,z)$, 
$I^{\hf,0}(\ast,z)$ and $I^{\hf,\hf}(\ast,z)$ are all 
integers and those of $z$ in $I^{\hf,\st}(\ast,z)$ are half-integers, 
that is, in $\hf+{\bf Z}$. \\
(2) $I^W(\ast,z)$ satisfies the derivation.\\
(3) $I^W(\ast,z)$ satisfies the supercommutativity:
$$\begin{array}{l}
I^W(v,z_1)I^W(v',z_2) \sim I^W(v',z_2)I^W(v,z_1), \cr 
I^W(v,z_1)I^W(u,z_2) \sim I^W(u,z_2)I^W(v,z_1), \cr
I^W(u,z_1)I^W(u',z_2) \sim (-1)I^W(u',z_2)I^W(u,z_1), 
\end{array} \eqno{(4.15)} $$
for $v,v'\in M_{\bar{0}}$ and $u,u'\in M_{\bar{1}}$. 
Here $ A(z_1,z_2)\sim B(z_1,z_2)$ implies that
 $$ (z_1-z_2)^NA(z_1,z_2)=(z_1-z_2)^NB(z_1,z_2) $$
for sufficiently large integer $N$, 
see Proposition 4.3.2 in {\rm \cite{FLM}}.   
In particular, $I^{\ast,\st}(\ast,z)$ satisfies 
the supercommutativity: \\
$$\begin{array}{l}
I^{0,\st}(v,z_1)I^{0,\st}(v',z_2) 
\sim I^{0,\st}(v',z_2)I^{0,\st}(v,z_1), \cr 
I^{0,\st}(v,z_1)I^{\hf,\st}(u,z_2) 
\sim I^{\hf,\st}(u,z_2)I^{0,\st}(v,z_1), \cr 
I^{\hf,\st}(u,z_1)I^{\hf,\st}(u',z_2) 
\sim -I^{\hf,\st}(u',z_2)I^{\hf,\st}(u,z_1), 
\end{array}  \eqno{(4.16)}$$
for $v,v'\in M_{\bar{0}}$ and $u,u'\in M_{\bar{1}}$. 
\end{rmk}

We next show an associativity. 
 
\begin{rmk}
Clearly,  $V_{2{\bf Z}x}=\{L(\hf,0)\otimes L(\hf,0)\}\ \oplus\ \{ L(\hf,\hf)
\otimes L(\hf,\hf)\}$ and it becomes a vertex operator algebra and 
$W=V_{\hf x+2{\bf Z}x}$ is a $V_{2{\bf Z}x}$-module as we showed. Viewing 
$L(\hf,0)\otimes L(\hf,0)$-module,  the module vertex 
operator $Y^W$ has the following structure by (4.14).   
$$ (I\otimes I)(u\otimes v,z)=
\left\{ 
\begin{array}{ll}
I^{0,\st}(u,z)\otimes I^{0,\st}(v,z)  &\mbox{for } u,v\in L(\hf,0), \cr
I^{\hf,\st}(u,z)\otimes I^{\hf,\st}(v,z)  &\mbox{for } u,v\in L(\hf,\hf). 
\end{array} \right. \eqno{(4.17)}$$
In particular, 
$I\otimes I$ satisfy the associativity and the derivation. \\
For example, for $u,v, u',v'\in L(\hf,\hf)$, 
$$ \begin{array}{l}
(I^{0,\st}\otimes I^{0,\st})\left((u\otimes v)_n(u'\otimes v'), z\right) \cr
=\left( (I^{\hf,\st}\otimes I^{\hf,\st})(u\otimes v,z)\right)_n 
\left((I^{\hf,\st}\otimes I^{\hf,\st})(u'\otimes v',z)\right). 
\end{array} $$
It is known that $Y(\ast,z)$ satisfies :
$$  Y(x(n)v, z)=Y(x(-1){\bf 1},z)_nY(v,z).  $$
Namely, $Y(\ast,z)$ satisfies the associativity with the actions of 
$x(-1){\bf 1}\in L(\hf,\hf)\otimes L(\hf,\hf)$.  
By the direct calculation, we see that 
the normal product satisfies the associativity:
$$  (a(z)_nb(z))_mc(z)=\sum_{i=0}^{\infty}(-1)^i\pmatrix{n\cr i}
(a(z)_{n-i}b(z)_{m+i}c(z)-(-1)^nb(z)_{n+m-i}a(z)_ic(z).  $$
Since $Y(\ast,z)$ satisfies the associativity with the actions of 
$L(\hf,0)\otimes L(\hf,0)$, $Y(\ast,z)$ satisfies the associativity with 
the actions of $L(\hf,\hf)\otimes L(\hf,\hf)$.  
Therefore, for 
$u^1\otimes u^2\in L(\hf,\hf)\otimes L(\hf,\hf)$ and
$u^3\otimes v\in L(\hf,\hf)\otimes L(\hf,0)$, we calculate on 
the actions on $V_{(\hf+{\bf Z})x}$ as follows: 
$$ \begin{array}{rl}
&Y((u^1\otimes u^2)_n(u^3\otimes v),z) \cr
=&Y(u^1\otimes u^2,z)_nY(u^3\otimes v,z) \cr
=&Res_{x}\{ (x-z)^nY(u^1\otimes u^2,x)Y(u^3\otimes v,z)
-(-z+x)^nY(u^3\otimes v,z)Y(u^1\otimes u^2,x) \} \cr
=&Res_{x}\{ (x-z)^n
\pmatrix{1&0\cr 0&-1}\otimes (I\otimes I)(u^1\otimes u^2,x)
\pmatrix{0&1\cr 1&0}\otimes (I\otimes I)(u^3\otimes v,z)\cr 
&-(-z+x)^n
\pmatrix{0&1\cr 1&0}\otimes (I\otimes I)(u^3\otimes v,z)
\pmatrix{1&0\cr 0&-1}\otimes (I\otimes I)(u^1\otimes u^2,x) \} \cr
=&\pmatrix{0&1\cr -1&0}
Res_{x}\{ (x-z)^n
\otimes I\otimes I(u^1\otimes u^2,x)
\otimes I\otimes I(u^3\otimes v,z) \cr
&+(-z+x)^n
\otimes I\otimes I(u^3\otimes v,z)
\otimes I\otimes I(u^1\otimes u^2,x) \}. 
\end{array} $$
On the other hand, since 
$(u^1\otimes u^2)_n(u^3\otimes v)\in L(\hf,0)\otimes L(\hf,\hf)$, 
we get
$$ \begin{array}{rl}
&Y((u^1\otimes u^2)_n(u^3\otimes v),z) \cr
=&\pmatrix{0 &1 \cr -1& 0}
\otimes(I\otimes I)((u^1\otimes u^2)_n(u^3\otimes v),z). 
\end{array} $$

Since $I^{\ast,\st}(\ast,z)$ satisfies the 
associativity with the actions of $L(\hf,0)$, we obtain :
$$ I^{0,\st}(u_nv,z)=I^{\hf,\st}(u,z)_nI^{\hf,\st}(v,z)  $$
for $u,v\in L(\hf,\hf)$, where 
the normal product of $I^{\hf,\st}(\ast,z)$ and $I^{\hf,\st}(\ast,z)$  
is given by 
$$ a(z)_nb(z)=Res_x\{ (x-z)^na(x)b(z)+(-z+x)^nb(z)a(x) \}  $$
and we will call this normal product "super normal product" and 
the above associativity "superassociativity". 
Namely, $I^{h,\st}(\ast,z)$ and $I^{k,\st}(\ast,z)$ satisfies the 
superassociativity:
$$ \begin{array}{l}
 I^{h+k,\st}(u_nv,z)\cr
 =Res_x\{ (x-z)^nI^{h,\st}(u,x)I^{k,\st}(v,z)
-(-1)^{|u||v|}(-z+x)^nI^{k,\st}(v,z)I^{h,\st}(u,x) \}  
\end{array} \eqno{(4.18)} $$
for $u\in L(\hf,h), v\in L(\hf,k)$ and $h,k=0,\hf$, where 
$|v|=0$ if $v\in L(\hf,0)$ and $|v|=1$ if $v\in L(\hf,\hf)$.  
\end{rmk}

Let extend it into the tensor product of many $M_{\bar{0}}$ and $M_{\bar{1}}$. 
Using the tensor product and (4.18), we obtain the following: \\

\begin{prn}
For two binary words $(a^i), (b^i)$ of even length $n$ and 
$u\in \otimes_{i=1}^nL(\hf,{a^i\over 2})$, \\  
$v\in \otimes_{i=1}^nL(\hf,{b^i\over 2})$, $\otimes I$ satisfies 
the superassociativity and the derivation: 
$$\begin{array}{l}
(\otimes I^{(a^i+b^i)/2,\st})(u_nv,z)
=\left( (\otimes I^{a^i/2,\st})(u,z)\right)_n
\left( (\otimes I^{b^i/2,\st})(v,z)\right) \cr
 [L(-1), (\otimes I^{a^i/2,\st})(u,z)]
={d\over dz}\left\{ (\otimes I^{a^i/2,\st}(u,z)\right\}. 
\end{array} \eqno{(4.19)}$$
We often abuse the notation $0,1\in {\bf Z}_2$ to denote 
integers $0,1$. For example, $a_i/2=0$ for $a_i=0$ and $a_i/2=\hf$ 
for $a_i=1$. 
\end{prn}

\pr 
By reordering the coordinates, we may assume 
$(a^i)=(1^{2s}0^{n-2s})$.  Dividing the coordinates into a set of pairs 
$\{1,2\}\{3,4\}\cdots \{2s-1,2s\}\{2s+1,...,n\}$ and then applying the 
above results to the pair $\{i,i+1\}$, we have the 
superassociativity for each $(a^i,a^{i+1})=(1,1)$ and 
$(b^i,b^{i+1})$.  For $a_i=0$, we have the associativity. 
Taking the tensor product of them, 
we obtain the desired superassociativity.  
The derivation comes from Remark 4. 
\prend

\subsection{Tensor product of SVOA}
Let us start a construction of VOA from an even binary linear code $D$ 
using the tensor product of vertex operator superalgebras.

In the previous subsection, we constructed the vertex 
operator superalgebra (SVOA) 
$$  M=M_{\ol{0}}\ops M_{\ol{1}}=L(\hf,0)\ops L(\hf,\hf), \eqno{(4.20)} $$
where $M_{\ol{0}}=L(\hf,0)$ is the even part and 
$M_{\ol{1}}=L(\hf,\hf)$ is the odd 
part.  For $v\in M_{\ol{i}}$ and $i=0,1$, $|v|$ denotes $i$.
We note that the notion of vertex operator superalgebra 
is given by supercommutativity:
$$Y(v,z_1)Y(w,z_2) \sim (-1)^{|v||w|}Y(w,z_2)Y(v,z_1). \eqno{(4.21)} $$

The tensor product of SVOAs $(M^1,Y^1),...,(M^n,Y^n)$ is given 
by the following:  \\
The full Fock space is 
$$   \hat{M}_F=M^1\ots ...\ots M^n  $$
and we define the vertex operator $\hat{Y}$ of 
$v^1\ots...\ots v^n\in M^1\ots \cdots \ots M^n$ by 
$$  \hat{Y}(v^1\ots ...\ots v^n,z)
=Y^1(v^1,z)\ots ...\ots Y^n(v^n,z)   \eqno{(4.22)}$$
and extend it to $\hat{M}_F$ linearly.

In this paper, we will take copies of $M$ as $M^i$, that is, 
$M^1\cong \cdots \cong M^n\cong M=L(\hf,0)\ops L(\hf,\hf)$. \\
For a word $\de=(d_1,...,d_n)\in {\Z}_2^n$, set 
$$\hat{M}_{\de}=\otimes_{i=1}^n (M^i)_{\ol{d_i}}, \eqno{(4.23)} $$
where $(M^i)_{\ol{0}}$ and $(M^i)_{\ol{1}}$ are 
the even part and the odd part of $M^i\cong L(\hf,0)\ops L(\hf,\hf)$, 
respectively. For example, 
$\hat{M}_{(0\cdots 0)}\cong L(\hf,0)\ots \cdots \ots L(\hf,0)$ 
and $\hat{M}_{(1\cdots 1)}\cong L(\hf,\hf)\ots \cdots \ots L(\hf,\hf)$.

\vspace{3mm}

By the definition of $\hat{M}_{\de}$ and the ${\Z}_2$-gradations of SVOAs, 
we have the following: \\

\begin{lmm}  
Let $\de=(d_1,...,d_n)$ and $\ga=(g_1,...,g_n)$ be words. 
For $v\in \hat{M}_{\de}$ and $w\in \hat{M}_{\ga}$, we obtain 
 $$    v_mw\in \hat{M}_{\de+\ga} \quad \mbox{ for any }m\in {\Z}. 
 \eqno{(4.24)} $$
\end{lmm}

Using the supercommutativity $(4.15)$ in SVOA, 
we have the following lemma.\\
	
\begin{lmm}
Let $\de=(d_1,...,d_n)$ and $\ga=(g_1,...,g_n)$ be words. 
For $v\in \hat{M}_{\de}$ and $w\in \hat{M}_{\ga}$, 
 we obtain  
$$ \hat{Y}(v,z_1)\hat{Y}(w,z_2)\sim 
(-1)^{\langle\de,\ga\rangle}\hat{Y}(w,z_2)\hat{Y}(v,z_1). \eqno{(4.25)} $$
Here $\langle\de,\ga\rangle=d_1g_1+...+d_ng_n $.  

In particular, if $\delta$ has an even weight, then 
$\hat{Y}(v,z)$ satisfies the commutativity with 
$\hat{Y}(v,z)$ itself for $v\in \hat{M}_{\de}$.
\end{lmm}

\vspace{3mm}

For the purpose of constructing a VOA, we need the commutativity. 
We shall use a central extension $\hat{D}$ of $D$ by $\pm 1$ in order to 
modify the supercommutativity of the above vertex operators. 
Let $\nu_i$ denote the word $(0,...,0,1,0,...,0)$ whose $i$-th entry is one 
and the other entries are all 0 and let $e^{\nu_i}$ be 
formal elements satisfying 
$ e^{\nu_i}e^{\nu_i}=1$ and $e^{\nu_i}e^{\nu_j}=-e^{\nu_j}e^{\nu_i}$ 
for $i\not=j$.  For any even word $\al=\nu_{j_1}+\cdots +\nu_{j_k}$ 
with $j_1<\cdots <j_k$, set 
$$e^{\al}=e^{\nu_{j_1}}e^{\nu_{j_2}}\cdots e^{\nu_{j_k}}. \eqno{(4.26)}$$

It is easy to see the following result. \\

\begin{lmm}
For $\al,\be$, 
 $$ \left\{  \begin{array}{ll}
 e^{\al}e^{\be}=(-1)^{\langle\al,\be\rangle}
e^{\be}e^{\al} &\mbox{ if } |\al||\be| \mbox{ is even } \cr
 e^{\al}e^{\be}=-(-1)^{\langle\al,\be\rangle}
e^{\be}e^{\al} &\mbox{ if } |\al||\be| \mbox{ is odd } 
\end{array} \right. 
\eqno{(4.27)} $$
\end{lmm}

\pr
Since  $$e^{\al}e^{\nu_i}=
\left\{ \begin{array}{lll}
(-1)^{|\al|}e^{\nu_i}e^{\al} &\mbox{ if }& \langle\al,\nu_i\rangle=0 \cr
(-1)^{|\al|-1}e^{\nu_i}e^{\al} &\mbox{ if }& \langle\al,\nu_i\rangle=1, \cr
\end{array} \right. \eqno{(4.28)} $$
we have the desired results. 
\prend

In order to combine $(4.27)$ into the vertex operator $\hat{Y}$, set 
$$M_D=\oplus_{\de \in D}(\hat{M}_{\de}\ots e^{\de}) \eqno{(4.29)} $$
and define a new vertex operator $Y$ by
 $$  Y(v\ots e^{\be},z)=\hat{Y}(v,z)\ots{e^{\be}} \eqno{(4.30)} $$ 
for $v\ots e^{\be} \in \hat{M}_{\be}\ots e^{\be}$. 
We then obtain the desired commutativity: 
$$ Y(v,z_1)Y(w,z_2)\sim Y(w,z_2)Y(v,z_1). \eqno{(4.31)}$$
Let ${\bf w}^i$ be the Virasoro element of $M^i$. Then 
$$ {\bf w}
=\sum_{i=1}^n({\bf 1}^1\ots ...\ots {\bf 1}^{i-1}\ots {\bf w}^i\ots 
{\bf 1}^{i+1}\ots ...\ots {\bf 1}^n)\ots e^0  \eqno{(4.32)}$$
satisfies the desired properties of the Virasoro element of $M_D$ and 
$$ {\bf 1}={\bf 1}^1\ots ...\ots {\bf 1}^n)\ots e^0  $$
is the vacuum of $M_D$, where ${\bf 1}^i$ is the vacuum of $M^i$. 
Finally, we have the following theorem. \\

\begin{thm} 
If $D$ is an even binary linear code, then 
$(M_D, Y, {\bf w}, {\bf 1})$ 
is a VOA.
\end{thm}

By the choice of our cocycle, we easily obtain the following result: \\

\begin{thm}
 ${\rm Aut}(D)$ is an automorphism group of VOA $(M_D,Y)$.
\end{thm}

\pr
 ${\rm Aut}(D)$ acts on $\{ \hat{M}_{\al}\ :\  \al\in D\}$ as permutations 
and also commutes with the products $(4.28)$ 
in the central extension $\{\pm e^{\be}\ |\ \be\in D\}$. Hence it becomes 
an automorphism group of $(M_D,Y)$.  
\prend

\begin{dfn}
For a word $\be$, $M_D$ has an automorphism $\delta_{\be}$ given by 
$$\delta_{\be}:u^{\al}\to (-1)^{\langle \be,\al\rangle}u^{\al} \eqno{(4.36)}$$
for $u^{\al}\in M_{\al}$. 
This is an automorphism induced from the inner automorphism 
of $M=M_{\bar{0}}\oplus M_{\bar{1}}$ and    
we will call this an inner automorphism of $M_D$.  
\end{dfn}

\vsp

\noindent
{\bf Notation} \qquad 
We always fix a lowest weight vector $q$ of $M_{\ol{1}}=L(\hf,\hf)$ 
with $q_{-2}q=2{\bw}$, see (4.9) and (4.10)  
and $q^i$ denotes such $q$ of each $M^i$. 
For a word $\al=(a^1,...,a^n)$ of length $n$, set 
$$ q^{\al}=(\otimes_{i=1}^nq^{a^i}), \eqno{(4.33)} $$
where $q^{a^i}=q^i$ if $a_i=1$ and $q^{a^i}={\bf 1}^i$ if $a_i=0$. 
Then $q^{\al}\otimes e^{\al}$ is a lowest weight vector of $M_{\al}$. 
For a $T$-module $X=\otimes_{i=1}^n L(\hf,b^i)\otimes R$ with some formal 
vector space $R$, we define an 
intertwining operator $I(q^{\al}\otimes e^{\al},z)$ of 
$q^{\al}\otimes e^{\al}$ of type 
$\pmatrix{ \otimes L(\hf,h^i+a^i/2)\otimes e^{\al}R \cr 
\otimes L(\hf,a^i/2)\otimes e^{\al} \quad 
\otimes L(\hf,h^i)\otimes R }$ by 
$$ (\otimes I)(q^{\al}\otimes e^{\al},z)
=(\otimes I^{(h^i)})(q^{\al}\otimes e^{\al},z)
=\otimes_{i=1}^n (I^{a^i/2,h^i}(q^{a^i},z))\otimes e^{\al}, \eqno{(4.34)} $$ 
where $\al=(a^i)$ and $I^{1/2,h^i}(q^i,z)$ as in $(4.12)$ 
and $I^{0,h^i}({\bf 1}^i,z)=1$. Here $h^i+a^i/2$ denotes the fusion 
rules (1.1), that is, $0+0=\hf+\hf=0,0+\hf=\hf+0=\hf,\st+0=\st+\hf=\st$. 
By Remark 4, the powers of $z$ in $I^{\hf,\st}(u,z)$ are ${\hf}+{\Z}$ for 
$u\in M_{\bar{1}}$ and those of $z$ in the other intertwining operators are 
all integers. 
Hence, for an even word $\al=(a^i)$ and a $T$-module 
$X$ with the $\st$-word $\tilde{h}(X)$,  
the powers of $z$ in $(\otimes I)(q^{\al}\otimes e^{\al},z)$ are 
in ${\Z}+\hf\langle \al,\tilde{h}(X)\rangle $, that is,
$$ (\otimes I)(q^{\al}\otimes e^{\al},z)
\in {\rm Hom}(X,M_{\al}\times X)[[z,z^{-1}]]
z^{\hf\langle \al,\tilde{h}(X)\rangle}. \eqno{(4.35)}  $$ 
\owari

We have thought of even code. 
We will next treat words of odd weights, too.  
By the direct calculation, we have the following. \\

\begin{lmm}
For $u\in M_{\de}$, $v\in M_{\ga}$, 
$Y(u,z_1)Y(v,z_2)\sim (-1)^{|\de||\ga|}Y(v,z_2)Y(u,z_1), $
where $|\de|$ and $|\ga|$ denote the weights of $\de$ and $\ga$, 
respectively. 
\end{lmm}

Namely, we have the following result: \\

\begin{thm}   
$(M_{{\Z}_2^n}, Y, {\bf w}, {\bf 1})$ is a SVOA. 
\end{thm}

\begin{rmk}
If $D$ is an even linear code of length $n$, then $M_D$ is a vertex operator 
subalgebra of $M_{{\Z}_2^n}$.  As $M_D$-modules, $M_{{\Z}_2^n}$ 
splits into a direct sum $\oplus_{\al\in {\Z}_2^n/D}M_{\al+D}$ 
of irreducible $M_D$-submodules $M_{\al+D}=\sum_{\be\in \al+D}M_{\be}$. 
\end{rmk}

\begin{dfn}
Since $M^i_{\ol{0}}\cong L(\hf,0)$ contains a conformal vector ${\bf w}^i$ 
with central charge $\half$ as the Virasoro element,  
$M_{(0,...,0)}=\otimes_{i=1}^n L(\hf,0)$ contains mutually orthogonal 
$n$ conformal vectors 
$$e^i={\1}^1\ots \cdots \ots {\1}^{i-1}\ots {{\bf w}^i}\ots 
{\1}^{i+1}\ots\cdots \ots {\1}^n. \eqno{(4.37)} $$
We will call the set $\{e^1,...,e^n\}$ ``the coordinate conformal 
vectors''. 
\end{dfn}

The most useful example in this paper 
is a VOA $M_{H_8}$ constructed from 
the $[8,4,4]$-Hamming code $H_8$, which we will denote 
by $V_{H8}$ and we will 
classify its representations in Sec.5.

\subsection{Uniqueness of tensor product construction}
Let $(V,Y)$ be a simple VOA with a set $\{e^i \ |\ i=1,\cdots, n\}$ of 
mutually orthogonal rational conformal vectors 
such that the sum $\sum_{i=1}^n e^i$ 
is the Virasoro element ${\bf w}$. 
Set 
$$T=<\!e^1,...,e^n\!>\cong \ots_{i=1}^n L(\hf,0).   $$ 
Assume that there is no $\st$ entry in the set ${\rm lwr}(T,V)$ of 
lowest weight rows $(h^1,...,h^n)$ of $V$.  

We will prove the following result: \\

\begin{thm}
Under the above assumptions, $(V,Y)$ 
is isomorphic to $M_D$ for some 
even linear binary code $D$. 
\end{thm}

\pr
By our assumption, every $e^i$ is a conformal vector 
of type two and defines an 
automorphism: 
$$\sigma_{e_i}\ :\  \left\{ 
\begin{array}{rll}
1 &\mbox{on }U \cong L(\hf,0) &\mbox{ as $<\!e_i\!>$-modules,} \\
-1 &\mbox{on }U \cong L(\hf,\hf) &\mbox{ as $<\!e_i\!>$-modules,}\\ 
\end{array} \right. \eqno{(4.38)} $$
by the second binary mode in $(1.2)$, see \cite{M1}.  
Set $R=<\!\sigma_{e_i}\ |\ i=1,...,n\!>$. Then $R$ 
is an elementary Abelian 2-group. Let $Irr(R)$ denote the 
set of all irreducible linear characters of $R$.  
Then we have the decomposition $V=\oplus_{\chi\in Irr(R)} V^{\chi}$ 
into the direct sum of 
irreducible $V^R$-modules $V^{\chi}$, where 
$V^{\chi}=\{v\in V\ | \ g(v)=\chi(g)v \mbox{ for all }g\in R\}$ is 
the weight space of $\chi$. Then the fixed 
point space 
$V^{R}$ is a direct sum of irreducible $T$-modules isomorphic to 
$\otimes L(\hf,0)$.  Since $\dim V_0=1$, we have $V^R\cong \otimes L(\hf,0)$ 
and so $V^R=T$. 
By the result in \cite{DM}, $V^{\chi}$ is an irreducible 
$T$-module and hence we have $V^{\chi}\cong M_{d(\chi)}$ as $T$-modules 
for some binary word $d(\chi)$. 
Let $D$ to be the set of all such words $d(\chi)$.
As they also proved by Lemma 3.1 in \cite{DM}, for two $\chi, \phi\in Irr(R)$, 
there are $u\in V_{\chi}, v\in V_{\phi}$ and $n\in {\Z}$ such that 
$u_nv\not=0$.  Hence, by the fusion rules for Ising models $(1.1)$, we 
have that $D$ is closed under the addition, that is, $D$ is a 
binary linear code.  Since the weight of elements in $V$ are all integers, 
$D$ is an even code and $ V\cong M_D $ as $T$-modules. 
Let us recall the fusion rules of irreducible modules of Ising model:
 $$ L(\hf,0) \mbox{ is an identity and } L(\hf,\hf)\ts L(\hf,\hf)=L(\hf,0). $$
These imply that the intertwining spaces 
$I\pmatrix{0 \cr \hf\quad \hf}$ and $I\pmatrix{\hf\cr \hf\quad 0}$
have one dimension.  Choose a fixed vector $q$ 
of $(M_{\ol{1}})_{\hf}$
so that $q_{-2}q=2{\bw}$, where ${\bw}$ is the Virasoro element of 
$M_{\ol{0}}=L(\hf,0)$. 
Let $I^{\hf,0}(q,z)$ and 
$I^{\hf,\hf}(q,z)$ 
be intertwining operators of $q$ of types $\pmatrix{L(\hf,0) \cr 
L(\hf,\hf) \quad L(\hf,\hf)}$ and $\pmatrix{L(\hf,\hf) \cr
L(\hf,\hf)\quad L(\hf,0)}$, respectively, such that 
$$ Y(q,z)=\pmatrix{ 0 &I^{\hf,\hf}(q,z) \cr I^{\hf,0}(q,z) & 0} 
\in {\rm End}(M)[[z,z^{-1}]]\eqno{(4.39)} $$
is the vertex operator of $q$ in the SVOA $M=M^0\oplus M^1$ as in $(4.9)$. 
For $\be=(b^1,...,b^n)\in D$, 
take a lowest weight vector 
$q^{\be}\otimes e^{\be}=(\otimes_i q^{b^i})\otimes e^{\be}$ of $M_{\be}$ 
as in (4.33). 
Since the intertwining space has one dimensional, the restriction:
$$Y(q^{\be}\otimes e^{\be},z)_{|M_{\xi}}\ :\ M_{\xi}\to M_{\xi+\be}
\eqno{(4.40)}$$ 
is a scalar times of 
$(I^{\xi})(q^{\be}\otimes e^{\be},z)
=(\otimes_i I^{b^i/2,\ast}(q^{b^i},z))\ots e^{\be}$ 
for $\be,\xi\in D$, say 
$$ Y(q^{\be}\otimes e^{\be},z)_{|M_{\xi}}
=\la(\be,\xi)I^{\xi}(q^{\be}\otimes e^{\be},z).   $$
Since this is true for any other elements, we have 
$$ Y(u^{\be}\otimes e^{\be},z)_{|M_{\xi}}
=\la(\be,\xi)I^{\xi}(u^{\be}\otimes e^{\be},z)   \eqno{(4.41)} $$
for any $u^{\be}\otimes e^{\be}\in M_{\be}$. 
The mutual commutativity of 
$\{ Y(u^{\be}\otimes e^{\be},z)\ |\ \be\in D\}$ imply 
$$ \la(\be,\al+\xi)\la(\al,\xi)
=\la(\al,\be+\xi)\la(\be,\xi) \eqno{(4.42)} $$
for any $\al,\be,\xi\in D$. 
Namely, $\la(\ast,\ast)$ is a 2-cocycle.  
Since $Y$ and $I^{\xi}$ satisfy the associativity, we have  
$$ \la(\be,\al+\xi)\la(\al,\xi)
=\la(\al,\be+\xi)\la(\be,\xi)=\la(\al+\be,\xi). \eqno{(4.43)}$$
In particular, $\la(0,\be)=1$ for any $\beta$ and $\la(\al,\be)=\pm 1$.  
Using the skew-symmetry, we have 
$\la(\be,0)=1$ for any $\beta$. 
Substituting $\xi=\al$ into $(4.43)$, we have 
$\la(\al,\al)=\la(\al+\be,\al)$ for any $\al,\be$. Hence 
$\la(\al,\be)=\la(\be,\be)=\la(0,\be)=1$ for any $\al, \be$. 
Namely, $Y$ is uniquely determined. 
\prend

\subsection{Examples}

Let $L_0={\Z}x^1+{\Z}x^2+\cdots {\Z}x^n$ 
be a lattice with $<x^i,x^j>=\delta_{i,j}$.  
Then $L=\{ \sum a^ix^i: \sum a^i\equiv 2 \pmod{2} \}$ is an even lattice. 
Let $V_L$ is the lattice VOA constructed by \cite{FLM}. 
$V_L$ contains $V_{2{\Z}x_1}\otimes \cdots \otimes V_{2{\Z}x_n}$ and hence  
$V_L$ contains a set 
$\{ e^1,e^2,\cdots, e^{2n-1},e^{2n}\} $ 
of mutually orthogonal $2n$ conformal vectors 
Set $T=<e^1,...,e^{2n}>$.  Since $V_L$ is a subspace of 
$V_{{\Z}x^1}\otimes \cdots \otimes V_{{\Z}x^n}$, 
there is no $\st$-entries in ${\rm lwr}(T,V_L)$.  
Let $D$ be the code consisting of all even words 
of length $2n$.  It is easy to see that $V_L$ is isomorphic to $M_D$.

\section{ $M_D$-modules}
\subsection{Classification}
In this subsection, we will classify the irreducible $M_D$-modules 
and study their fusion rules.

Let $\{e^1,...,e^n\}$ be the set of coordinate conformal 
vectors of $M_D$ and set $T=<\!e^1\!>\ots\cdots\ots<\!e^n\!> \cong 
\otimes_{i=1}^n L(\hf,0)$.  
We note that all irreducible $T$-submodules of $M_D$ 
has no $L(\hf,\st)$-entry. 
Let $(X,Y^X)$ be an irreducible $M_D$-module. Since $T$ is rational 
by Corollary 3.1,  we have a decomposition 
$X=\oplus U^i$ of $X$ as a direct sum of 
irreducible $T$-submodules $U^i$. 
By the fusion rules for Ising model, the $\st$-words of 
any irreducible $T$-submodules of $X$ are all the same, say 
$\tilde{h}(X)$. By Remark 4.(1), we have:

\begin{lmm}  $\tilde{h}(X)$ is orthogonal to $D$. 
\end{lmm}

Set $K=\{\al\in D|\al\subseteq \tilde{h}(X)\}$ and let $H$ be a maximal 
self-orthogonal subcode of $K$. 
Since $K$ is an even code, we have $H^{\perp}\cap K\subseteq H$. 
Let $\hat{K}=\{ \pm e^k:k\in K\}$ denote 
the central extension group of $K$ by $\pm 1$ induced from 
the inner products as in (4.26) and (4.27), then 
$\{\pm e^{\al}:\al\in H\}$ is 
a maximal normal Abelian subgroup of $\hat{K}$. Our aim in this 
subsection is to show that every irreducible $M_H$-submodule of $X$ determines 
the $M_D$-module structure on $X$ uniquely. 

\begin{thm}  
Let $(X,Y^X)$ be an irreducible $M_D$-module and 
$\{ X^i:i=1,...,k\}$ the set of non-isomorphic irreducible 
$T$-submodules of $X$.  Then there are irreducible $\hat{K}$-modules 
$Q^i$ on which $-e^0\in \hat{K}$ acts as $-1$ such that 
$X\cong \oplus_{i=1}^k (Q^i\otimes X^i)$ as $M_K$-modules.  
\end{thm}

\pr
Let $U$ be a homogeneous component of $X$ generated by all $T$-submodules 
isomorphic to $X^1$ and 
$U=\oplus_{i=1}^k U^i$ the decomposition of $U$ into a 
direct sum of irreducible $T$-submodules $U^i$.  
For $\al=(a^i)\in K$, set $q^{\al}$ as in $(4.33)$. 
By the fusion rules for Ising module $(1.1)$, 
$Y^X(q^{\al}\otimes e^{\al},z)U\subset U[[z,z^{-1}]]$ for $\al\in K$.  
Since the both of the tensor products of $L(\hf,0)$ and $L(\hf,\hf)$ with 
any irreducible $L(\hf,0)$-modules are irreducible, 
the vertex operator $Y^X(u^{\al}\otimes e^{\al},z)_{|U}$ 
of $u^{\al}\otimes e^{\al}$ has an expression 
$$A(e^{\al})\otimes \left( (\otimes I)(u^{\al},z)\right),$$ 
for $u^{\al}\otimes e^{\al}\in M_{\al}$.
where $A(e^{\al})$ is 
a $k\times k$-matrix and $(\otimes I)(\ast,z)$ is 
the tensor product of the intertwining operators in (4.12) and (4.13). 
Since $Y^X(u^{\al},z)$ satisfies the commutativity and 
$(\otimes I)(u^{\al},z)$ satisfies the super-commutativity, we see 
the supercommutativity : \\
$$  A(\al)A(\be)=(-1)^{\langle \al,\be\rangle}A(\be)A(\al). $$
Furthermore, since the both $Y^X(\ast,z)$ satisfies the associativity and 
$(\otimes I)(u^{\al},z)$ satisfy 
the superassociativity, we see the associativity : 
$$A(e^{\al})A(e^{\be})=A(e^{\al}e^{\be}) $$ and 
$A(\al)A(\al)$ is the identity matrix for all $\al, \be\in K$.  
This implies that $A$ is a matrix representation of the central 
extension $\hat{K}$ and $U=Q^1\otimes X^1$ for a $\hat{K}$-module $Q^1$.  
Suppose that $Q^1$ is not an irreducible $\hat{K}$-module and 
$Q^0$ is a proper submodule.  
Let $W$ be a subspace spanned by $\{ v_nw: v\in M_D, w\in Q^0, n\in {\Z}\}$. 
By Proposition 4.1 in \cite{DM}, we have $X=W$. 
On the other hand, by the fusion rules for Ising model $(1.1)$, 
$M_{\be+K}\times Q^0\otimes X^1$ does not contains a submodule 
isomorphic to $X^1$ for $\be\not\in K$ and so 
$W\cap U=Q^0\otimes X^1$, a contradiction.  
Hence, $Q^1$ is an irreducible $\hat{K}$-module on which 
$-e^0$ acts as $-1$.
\prend

As a corollary, we have the following:

\begin{cry}  If $E$ is a self-orthogonal binary linear code and 
$X$ is an irreducible $M_E$-module, then 
the multiplicities of irreducible $T$-submodules of $X$ are at most one.
In particular, if $X$ contains a $T$-submodule 
$\otimes_{i=1}^n L(\hf,\st)$, 
then $X\cong \otimes_{i=1}^n L(\hf,\st)$ as $T$-modules. 
\end{cry}

We next prove that $M_D$ is rational by the same argument. 
Namely, $M_D$ has finitely 
many non-isomorphic irreducible modules and every module is a direct sum of 
irreducible modules.

\begin{prn}
$M_D$ is rational. 
\end{prn}

\pr
Let $W$ be an indecomposable $M_D$-module.  
We will show that $W$ is an irreducible $M_D$-module. 
Suppose false and let $U$ be a proper $M_D$-submodule 
of $W$.  Viewing $W$ as a $T$-module, 
$W$ is a direct sum  $W=\oplus W^i$ of irreducible $T$-modules and 
assume $W^1\subseteq U$.  
Since the action of $M_D$ does not change the $\st$-words, 
every $W^i$ has the same $\st$-word $\be$ and the set 
${\rm lwr}(T,W)$ of lowest weight rows is an orbit of $D$. 
Set $K=\{ \al\in D: \al\subseteq \be\}$ and let $S$ be a direct sum 
of all $W^i\cong W^1$.  By the above theorem, $S\cong Q\otimes W^1$ 
and $S\cap U\cong Q^0\otimes W^1$ 
for a $\hat{K}$-module $Q$ and its submodule $Q^0$. 
Since the ${\bf C}\hat{K}$-modules are 
completely reducible, $Q$ is a direct sum $Q^0\oplus Q^1$ 
for some $\hat{K}$-module $Q^1$. 
By the assumption, we have $Q^1\not= 0$. 
Let $U^i$ be the space spanned by $\{ v_nw: v\in M_D, w\in 
Q^i\otimes W^1, n\in {\Z} \}$ for $i=0,1$. 
By Proposition 4.1 in \cite{DM}, they are $M_D$-modules. 
Then by the fusion rules, $U^0\cap U^1\cap Q\otimes W^1=0$. 
Namely, $U^0\cap U^1$ does not contain $T$-submodules isomorphic to $W^1$. 
Since the set $lwr(T,W)$ is an orbit of $D$, we obtain $U^0\cap U^1=0$ and 
so $W=U^0\oplus U^1$, which contradicts to the assumption. 
\prend

\subsection{Induced module}
We next prove that $M_D$-modules containing a fixed irreducible 
$M_K$-module are uniquely determined.  
Let's construct such an $M_D$-module. 

Since $\hat{K}$ is a direct product of an Abelian group and an 
extra-special 2-group, it is known that every irreducible $\hat{K}$ 
representation $\phi$ with $\phi(-e^0)=-1$ is induced from a linear 
representation $\chi$ of a maximal Abelian normal subgroup $\hat{H}$.  

Let $\be\in D^{\perp}$ and 
$K=\{\al=(a^i)\in D\ |\al \subseteq \be\}$. 
Let $H$ be a self-orthogonal subcode of $K_{\be}$ and 
choose any linear irreducible 
character $\chi:\pm \hat{H}\to \pm 1$ with $\chi(-e^0)=-1$.  
Let $F_{\chi}$ be a one dimensional $\hat{H}$-module such that 
$e^{\al}p=\chi(e^\al)p$ for $p\in F_{\chi}, \al\in H$. 
Take any $h^i=0,\hf,\st$ such that $\st$-word of $(h^i)$ is $\be$. 
Then $U=\otimes L(\hf,h^i)\otimes F_{\chi}$ is an $T$-module.
Define the module vertex operator $Y^U((\otimes u^i)\otimes e^{\al},z)$ 
for $u^i\in M{\ol{a^i}}$ on $U$ and $(a^i)\in H$ by 
$$(\otimes_{i=1}^n I^{a^i/2,h^i}(u^i,z))\otimes \chi(e^{\al}) \eqno{(5.1)} $$ 
and extend it to all elements in $M_H$ linearly. 
By Remark 5 and the associativity (4.19), we have:

\begin{lmm}  $Y^U(v,z)$ are all well defined for $v\in M_H$ 
and they satisfies the derivation, 
the associativity and the mutually commutativity. 
In particular, $U$ is an $M_H$-module. 
\end{lmm}

We denote the above $M_H$-module by $U((h^i))\otimes F$. 
Let $D$ be an even linear binary code and assume that 
$D$ is orthogonal to $\tilde{h}(U)$. 
We will define an induced $M_D$-module 
$X={\rm Ind}_{M_H}^{M_D}(U(h^i),\chi)$ from 
$U((h^i),\chi)$ as follows:   \\
Let $\{ \be^1=(b^1_i),\ldots,\be^s=(b^s_i)\}$ be 
 a transversal of 
$H$ in $D$, then $\{e^{\be^i}:i=1,\ldots,s\}$ is a transversal of 
$\hat{H}$ in $\hat{D}$.  Set 
$$ X=\oplus_{(\be^i)\in D/H}
 \{ U(h^i+{b^i\over 2})\otimes (e^{\be^i}\otimes_{\hat{H}} F_{\chi}) \}, 
 \eqno{(5.2)} $$
where $h^i+{b^i\over 2}$ denotes the fusion rules for Ising models. 
Since $e^{\be+h}\otimes_{\hat{H}} F_{\chi}=e^{\be}\otimes F_{\chi}$ 
for $h\in H$ 
and $(h^i+{b^i\over 2})=(h^i)$ for $(b^i)\in H$, 
$X$ does not depend on the choice of transversal of $H$ in $D$ and 
$X$ becomes an $M_H$-module. 

We define the module vertex operator $Y(u^{\ga}\otimes e^{\ga},z)$ 
of $u^{\ga}\otimes e^{\ga}$ by 
$(\otimes I)(u^{\ga},z)$ on the first term and $e^{\ga}$ on the second 
term for $\ga\in D$, where $(\otimes I)(u^{\ga},z)$ is as in $(4.34)$.
We denote this module by $Ind_{M_H}^{M_D}(U)$. 
We will show that these actions are all well defined and 
they satisfy the commutativity, the derivation and the associativity. 
Namely, we will prove the following:

\begin{prn}  $Ind_{M_H}^{M_D}(U)$ is an $M_D$-module. 
\end{prn}

\pr
First, we assume $\tilde{h}(U)=(1^n)$.  Then, $X=\oplus U((\st)_i)\otimes 
Q$, where $Q$ is a $\hat{K}$-module induced from $F$. 
Since $(\otimes I)(u^{\al},z)$ satisfies the derivation, the 
supercommutativity (4.16) and the superassociativity (4.19), 
$Y(u^{\al}\otimes e^{\al},z)$ satisfies the derivation,  
the commutativity and the associativity. Hence, $X$ is an $M_D$-module. 
We next assume that $\tilde{h}(U)=(0^n)$.  
Then, $H=\{ (0^n)\}$ and $F=F_{\chi}$ is a trivial module and 
$X=\oplus U(h^i+b^i/2)\otimes e^{\be^i}$ is isomorphic to 
$M_{(h^i)+D}$.  Hence, $X$ is also an $M_D$-module.  
We next treat the general case.  By the permutation of coordinates, 
we may assume $\tilde{h}(U)=(0^s1^t)$.  Let $S_n$ be the set of 
all even words of length $n$.  Since $D$ is orthogonal to $(0^s1^t)$, 
$D$ is a subcode of $S_s\oplus S_t$.  Divide the coordinates into 
the first $s$ coordinates and the last $t$ coordinates and set 
$H=H^0\oplus H^1$, where $H^0=\{(0^s)\}$ and 
$H^1=\{(a^{s+1},...,a^{s+t}):(a^i)\in H\}$.  
Clearly, $U\cong T^s\otimes U^1$ as $M_{H^0}\otimes M_{H^1}$-modules, 
where $T^s\cong \otimes_{i=1}^s L(\hf,h^i)$ is an $M_{H^0}$-module and 
$U^1\cong \otimes_{i=s+1}^{s+t}L(\hf,\st)$ is an $M_{H^1}$-module. 
Clearly, we see 
$$Ind_{M_H}^{M_D}(U)\subseteq   Ind_{M_H}^{M_{S_s\oplus S_t}}(U)
\subseteq Ind_{M_{H^0}}^{M_{S_s}}(T^s)\otimes Ind_{M_{H^1}}^{M_{S_t}}(U^1). $$
As we showed, $Ind_{M_{H^0}}^{M_{S_s}}(T^s)$ is an $M_{S_s}$-module and 
$Ind_{M_{H^1}}^{M_{S_t}}(U^1)$ is an $M_{S_t}$-module. We hence 
have an $M_{S_s\oplus S_t}$-module $Ind_{M_H}^{M_{S_s\oplus S_t}}(U)$, 
which is also an $M_D$-module containing  \\
$Ind_{M_H}^{M_D}(U)$.
Therefore, the module vertex operators satisfy the commutativity, the 
derivations, and the associativity.  Since $Ind_{M_H}^{M_D}(U)$ is 
invariant under the actions of $M_D$, we have the desired results. 
\prend

Since $q^{\ga}$ for $\ga\in D-H$ acts regularly on the set of 
irreducible $T$-modules $U$ with a $\st$-word $\be$, 
we have that if $M_D$-module $X$ contains 
a $M_H$-module $U$, then $X$ has to have the above structure on it. 
Therefore, we have proved the following theorem. \\

\begin{thm}
Let $X$ be an irreducible $M_D$-module with a $\st$-word $\be$. 
Set $K=\{\al\in D| \al\subseteq \be\}$ and let $H$ be a 
maximal self-orthogonal subcode of $K$. 
Then there is a pair 
$((h^i),\chi)$ of a lowest weight row $h=(h^1,...,h^n)$ 
and a linear character $\chi$ of $H$ such that 
$X=Ind_{M_H}^{M_D}(U((h^i),\chi)))$.
\end{thm}

As corollaries of the above theorem, we have the following results:

\begin{thm}
Let $X$ be an $M_D$-module with a $\st$-word $\tilde{h}$. 
Set $K=\{\al\in D:\al\subseteq \tilde{h}\}$ and let $H$ be a maximal 
self-orthogonal subcode of $K$.  Then the $M_D$-module structure on $X$ 
is uniquely determined by an $M_H$-submodule of $X$.  
\end{thm}

\begin{cry}  Let $D$ be an even binary linear code containing a 
self-dual subcode $E$.  
Assume that $X$ is an irreducible $M_D$-module with a $\st$-word $\be$. Set $K=\{\al\in D|\al\subseteq \be\}$ 
and $H=\{\al\in K|\langle \al,\be\rangle=0 \mbox{ for all }\be\in K\cap E\}$, 
then $X=\sum M^i\otimes T^i$ as $M_E$-modules, where 
$M^i$ are non-isomorphic $M_E$-modules and $T^i$ are 
irreducible $\hat{H}$-modules. 
In particular, the multiplicity of $M^i$ in $X$ is $\sqrt{|E+H:E|}$. 
We note that the actions of $M_E$ and $M_{H}$ are not always commutative.  
\end{cry}

\pr
Let $H'$ be a maximal self-orthogonal subcode of $K$ containing 
$K\cap E$.  
By the above theorem, 
there is an irreducible $M_{H'}$-module 
$U=\otimes L(\hf,h^i)$ such that 
$X=Ind_{M_{H'}}^{M_D}(U)$.  Clearly, $U$ is irreducible as $T$-module 
and $T\subseteq M_E$.  Let $M^1$ be a 
irreducible $M_E$-submodule of $X$ containing $U$. Let $M$ denote the 
$M_E$-submodule generated by all irreducible $M_E$-submodule isomorphic to 
$M^1$. Then by Theorem 5.2, we obtain 
$$M^1=\sum_{\be^j\in E/H\cap E}U(h^i+b^j_i)\otimes e^{\be^j}\otimes F.$$ 
Clearly, as $T$-submodules, we have 
$$ \begin{array}{l}
M\subseteq \sum_{\be^j\in (E+K)/H'}U(h^i+b^j_i)\otimes e^{\be^j}\otimes F\cr
=\sum_{\ga^j\in (K+E)/(E+H')}(\sum_{\be^k\in (E+H')/H'}U
(h^i+b^j_i+\ga^k_i)\otimes e^{\be^k})e^{\ga^j}\otimes F \cr
=\sum_{\ga^j\in (K+E)/(E+H')}(\sum_{\be^k\in E/(E\cap H')}U(h^i+b^j_i)
\otimes e^{\be^k})e^{\ga^j}\otimes F \cr
=\sum_{\ga^j\in (K+E)/(E+H')}M^1e^{\ga^j}\otimes F.  
\end{array}$$
It follows from direct calculation that 
$M^1e^{\ga^j}\otimes F$ is isomorphic to $M^1$ if and only if 
$\ga^j\in K$ and $\langle \ga^j,\be\rangle=0 \mbox{ for all }\be\in E\cap K$.
We hence have the desired decomposition and multiplicity.
\prend

\subsection{Fusion rule of $M_D$-modules}
Let us calculate the fusion rule of $M_D$-modules in this subsection. 
Restricting Proposition 11.9 in \cite{DL} into our case, we have:

\begin{thm}   Let $E$ be a subcode of $D$. 
Let $W^1, W^2, W^3$ be irreducible $M_D$-module 
and $U^1, U^2$ irreducible $M_E$-submodules of $W^1$ and $W^2$, respectively, 
then there is an injection map:
$$ \phi\ :\ I_{M_D}\pmatrix{W^3 \cr W^1 \quad W^2}
\to I_{M_E}\pmatrix{W^3\cr U^1\quad U^2}. $$
\end{thm}

\pr
For $I(\ast,z)\in I_{M_D}\pmatrix{W^3 \cr W^1 \quad W^2}$ and $v\in U^1$, \\
$I'(v,z)=I(v,z)_{|U^2}\in {\rm Hom}(U^2,W^3)\{z,z^{-1}\}$. 
Set $\phi(I(\ast,z))=I'(\ast,z)$. Clearly, 
$I'(\ast,z)\in I_{M_E}\pmatrix{W^3\cr U^2\quad U^2}$ and hence 
we have a map  
$$\phi: I_{M_D}\pmatrix{W^3 \cr W^1 \quad W^2} \to 
I_{M_E}\pmatrix{W^3 \cr U^1 \quad U^2}. $$
Clearly, $\phi$ is a linear map.  
We will show that $\phi$ is injective.  Namely, we will prove that 
if $I(v,z)_{|U^2}=0$ for all $v\in U^1$, then $I(\ast,z)=0$.  
By the commutativity of intertwining operators, we see 
$0=(z_1-z_2)^NY^3(u,z_1)I(v,z_2)w=(z_1-z_2)^NI(v,z_2)Y^2(u,z_1)w $
for all $u\in M_D$ and $w\in U^2$.  
Suppose $I(v,z_2)Y^2(u,z_1)w\not=0$, there is an integer $r$ such that 
$I(v,z_2)u^2_{r-1}w\not=0$ and 
$I(v,z_2)Y(u,z_1)w=I(v,z_2)(\sum_{i=0}(u^2_{-i+r-1}w)z_1^{i-r})$. 
However, since  
$(z_1-z_2)^NI(v,z_2)z_1^rY(u,z_1)w=0$ for a sufficiently large integer $N$, 
we obtain  
$(-z_2)^NI(v,z_2)u_{r-1}w=0$ by substituting $0$ into $z_1$ 
and so we have $I(v,z_2)u_{r-1}w=0$, which 
contradicts to the choice of $r$.   
Hence, $I(v,z_2)Y^2(u,z_1)w=0$.  Since the set $\{ u_mw:m\in {\Z}, u\in M_D, 
w\in U^2\}$  spans $W^2$ by Proposition 4.1 in \cite{DM}, 
we have $I(v,z_2)w=0$ for all $w\in W^2$. 
For $u \in M_D$, $v\in U^1$, $w\in W^2$ and $n\in {\Z}$, 
$$ I(u_nv,z)u=Res_{z_1}\{(z_1-z)^nY^3(u,z_1)I(v,z)w
-(-z+z_1)^nI(v,z)Y^2(u,z_1)w \}=0. $$ 
Hence, it follows from the associativity of intertwining operators that 
$I(v,z)=0$ for all $v\in W^1$. 
\prend

As a corollary of the above theorem, we will prove the 
following fusion rule. 

\begin{thm}
Let $X$ be an irreducible $M_D$-module with a $\st$-word $\tilde{h}(X)$, 
then the tensor product 
$$ M_{\al+D} \times X  $$
is an irreducible $M_D$-module.  Furthermore, if 
$\tilde{h}(X)$ is orthogonal to $\al$, 
then the intertwining operators $I(v,z)$ of type 
$\pmatrix{M_{\al+D}\times X \cr M_{\al+D}\ X}$ are in 
${\rm Hom}(X,M_{\al+D}\times X)[[z,z^{-1}]]$ for 
all $v\in M_{\al+D}$.  Namely, the 
the powers $z$ in $I(v,z)$ are all integers. 
If $\tilde{h}(X)$ is not orthogonal to $\al$, then 
$I(v,z)\in {\rm Hom}(X,M_{\al+D}\times X)[[z,z^{-1}]]z^{\hf}$. 
\end{thm}

\pr
Set $\al=(a^i)$ and $K=\{ \be\in D|\be\subseteq \tilde{h}(X)\}$.   
Let $U$ be an irreducible $M_D$-module 
such that \\
$I_{M_D}\pmatrix{U \cr M_{\al+D} \quad X}\not= 0$.  
By Theorem 5.1, we have a decomposition 
$X=\oplus_{\be=(b^i)\in D/K} (X^{\be}\otimes Q^{\be})$   
as a direct sum of irreducible $M_K$-modules $Q^{\be}\otimes X^{\be}$, 
where $Q^{\be}$ are irreducible $\hat{K}$-modules and 
$X^{\be}=\otimes L(\hf,h^i+b^i/2)$ are irreducible $T$-modules for 
some $h=(h^i)$. 
Similarly, we have $U=\oplus (Q^{\ga}\otimes U^{\ga})$ with  
irreducible $\hat{K}$-modules $Q^{\ga}$ and irreducible $T$-modules 
$U^{\ga}=\otimes L(\hf,h^i+b^i/2+a^i/2)$.  
By Theorem 5.4, we obtain  
$$ \dim I_{M_D}\pmatrix{U \cr M_{\al+D}\quad X} 
\leq \dim I_{M_K}\pmatrix{Q^{\be}\otimes X^{\be} \cr 
M_{\al+K}\quad Q^{\ga}\otimes U^{\ga}}, $$ 
where $\ga=\alpha+\be$.  
For $q^{\al}\otimes e^{\al}\in M_{\al+K}$, the intertwining operator 
$I(q^{\al}\otimes e^{\al},z)$ 
is expressed by 
$$ A \otimes \left( ( \otimes I)(q^{\al},z)\otimes e^{\al} \right)  $$ 
for some $k\times k$-matrix $A$. 
Moreover, the powers of $z$ in $I(q^{\al}\otimes e^{\al},z)$ are in 
${\Z}+\half\langle \tilde{h},\al\rangle$ by $(4.35)$. 
It follows from the commutativity of intertwining operators that 
$A$ satisfies the relation 
$$A\phi(s)=\psi(s)A  \eqno{(5.4)}$$ 
for all $s\in \hat{K}$, 
where $\phi$ and $\psi$ are the representations of $\hat{K}$ on 
$Q^{\be}$ and $Q^{\ga}$, respectively.  
Since $\phi$ and $\psi$ are irreducible, 
$A$ is uniquely determined up to the scalar multiple and so 
we have 
$\dim I_{M_D}\pmatrix{U \cr M_{\al+D}\quad X}\leq 1$.  
If $\dim I_{M_D}\pmatrix{U\cr M_{\al+D}\quad X}=1$, 
then as we showed the $\hat{K}$-module $Q^{\beta}$ is isomorphic to 
$Q^{\ga}$ and so the $M_D$-module structure on $U$ is uniquely 
determined by Theorem 5.3.
\prend

\section{Hamming code VOA $V_{H8}$}
\subsection{Definition of $V_{H8}$ and its twenty four conformal vectors}
An interesting property of the above construction is that 
the full automorphism group of $M_D$ has a normal subgroup which 
is a 3-transposition group. A 3-transposition group is a group 
generated by a conjugacy class of involutions 
and the product of any two involutions in this class has the order 
less than or equal to three. Especially, if 
$D$ contains a $[8,4,4]$-Hamming code $H_8$ as a subspace, 
it defines new sixteen rational
conformal vectors of central charge $\half$ and so we have new 
automorphisms given by conformal vectors (see \cite{M2}). \\

\begin{dfn}  An even binary linear code 
${\cal H}_8$ with the generator matrix 
$$ H=\left( 
\begin{array}{cccccccc}
 1 &  1 &  1 &  1 &  1 &  1 &  1 &  1 \cr
 0 &  0 &  0 &  0 &  1 &  1 &  1 &  1 \cr
 0 &  0 &  1 &  1 &  0 &  0 &  1 &  1 \cr
 1 &  0 &  1 &  0 &  1 &  0 &  1 &  0 \cr
\end{array} \right)   \eqno{(6.1)}$$
is called the $[8,4,4]$ Hamming code.
\end{dfn}

We can construct a VOA $M_{H_8}$ from 
$[8,4,4]$-Hamming code $H_8$ by the tensor products of SVOA $M$ and 
denote it by $V_{H8}$. Let $\{e^1,...,e^8\}$ be a set of coordinate 
conformal vectors. 

\vspace{3mm}

In ${\cal H}_8$, there are fourteen codewords of weight four. 
By the direct calculation, we can obtain the following: \\

\begin{lmm}  For a four-point set $\xi$, we write 
$u^{\xi}=q^{\xi}\otimes e^{\xi}$ in this section to simply the notation. 
Let $\al$ and $\beta$ be four-point sets. Then 
$u^{\al},u^{\be}\in (V_{H8})_2$ and we have:
 $$ \begin{array}{cl}
(u^{\al})_1(u^{\al})=2(\sum_{i\in \al}e^i) &  \cr
(u^{\al})_1u^{\be}=u^{\al+\be} & \mbox{ if }|\al\cap\be|=2 \cr
(u^{\al})_1u^{\be}=0 & \mbox{ if } \al\cap\be=\emptyset \cr
\langle e^i, e^j\rangle={1\over 4}\delta_{i,j} & \mbox{ and } \cr 
\langle u^{\al},u^{\be}\rangle=\delta_{\al,\be}. & 
\end{array}   \eqno{(6.2)}$$
\end{lmm}

As we showed in \cite{M2}, we can construct conformal vectors 
as follows: \\

\begin{thm}  In $V_{H8}$, we have the following conformal 
vectors with central charge $\half$: \\
 $$ s^{\al}={1\over 8}(e^1+\cdots +e^8)+{1\over 8}\sum_{\be\in C, \ |\be|=4}
(-1)^{\langle \al,\be\rangle}q^{\be}\otimes e^{\be} \eqno{(6.3)} $$
for a word $\al$.  
This is defined by the co-code ${\Z}_2^8/H_8$, that is, 
$s^{\al}=s^{\be}$ if and only if $\al-\be\in H_8$. 
In particular, we have sixteen new rational conformal vectors.  By 
the direct calculation using Lemma 6.1, we have 
$  \langle s^{\al},s^{\be}\rangle=0 \mbox{ if and only if }
\al+\be \mbox{ has even weight.} $
\end{thm} 

\vsp

Namely, we obtain the following three sets of eight mutually orthogonal 
conformal vectors of $V_{H8}$. 
$$ \{e^1,...,e^8\},
\qquad \{ s^{\be}\ :\quad |\be| \mbox{ odd weight }\},  
\qquad \{ s^{\al}\ :\quad |\al| \mbox{ even weight }\}. \eqno{(6.4)}$$

We can take $\{\nu_1=(10\cdots 0),\nu_2=(010\cdots 0),...,\nu_8=(0\cdots 01)\}$ 
as the set of representatives of the odd weight cosets of ${\Z}_2^8/H_8$ 
and $\{\nu_1+\nu_1=(0\cdots 0),\nu_1+\nu_2,...,\nu_1+\nu_8=(10\cdots 01)\}$ 
as that of even weight cosets of ${\Z}_2^8/H_8$.

\begin{dfn} We will use the following notation in this paper. \\
$$\begin{array}{lll}
 d^j=s^{\nu_1+\nu_j} &\mbox{ for }&j=1,...,8 \mbox{ and }\cr 
 f^i=s^{\nu_i} &\mbox{ for }&i=1,...,8.
\end{array} \eqno{(6.5)} $$
\end{dfn}

By the direct calculation using Lemma 6.1 and the results in \cite{M1}, 
we obtain the following lemma. \\

\begin{lmm}  There are exactly three sets of mutually orthogonal 
8 conformal vectors with central charge $\half$ in $V_{H8}$. 
\end{lmm} 

\pr
Clearly, since $H_8$ has no codeword of weight 2, $(V_{H8})_1=0$ 
by the definition of code VOA. 
Suppose false and let $\{g^1,...,g^8\}$ be another set of 8 conformal vectors.
Viewing $V_{H8}$ as a $<g^1,...,g^8>$-module, there is no $L(\hf,\st)$ 
since $\sum g^i$ is the Virasoro element and the weights of elements 
in $V_{H8}$ are integers. Therefore, $g^i$ are all of type 2 and so 
we have $<e^i,g^j>=1/32$. Since
$$(V_{H8})_2={\C}e^1+\cdots+{\C}e^8+\sum_{\be\in C, |\be|=4} {\C}u^{\be}, $$
if we set 
$g=\sum a^ie^i+\sum a^{\be}u^{\be}$, then we have $a_i=1/8$. 
Calculating $<g^i,g^i>$ and $g^i_1g^i$ by using (6.2), we have 
$\{g^1,...,g^8\}$ is equal to $\{f^1,...,f^8\}$ or $\{d^1,...,d^8\}$. 
\prend

\vsp

It is a routine work to see the action of $\sigma_{e_i}$ on $s^{\al}$.

\begin{lmm}  
$ \sigma_{e^i}(s^{\al})=s^{\al+\nu_i}. $
\end{lmm} 

\pr
$$ \begin{array}{l}
\sigma_{e^i}(s^{\al})=\sigma_{e^i}\left\{ {1\over 8}(\sum e^i)
+{1\over 8}(\sum (-1)^{<\al,\be>}u^{\be})\right\} \cr
={1\over 8}\sum e^i+\sum (-1)^{<\al,\be>}(-1)^{<\nu_i,\be>}u^{\be} \cr
={1\over 8}\sum e^i+\sum (-1)^{<\al+\nu_i,\be>}u^{\be} \cr
=s^{\nu_i+\al}.
\end{array} $$

\vsp

Namely, each involution $\sigma_{e_i}$ 
permutes the new sixteen conformal vectors regularly. 
We want to note one more thing. 
Since $[8,4,4]$-Hamming code is the only one 
self-dual doubly even code of length 8,  
$V_{H_8}$ is still isomorphic to $V_{H_8}$ as $<f^1,...,f^8>$-modules.

\subsection{Irreducible $V_{H8}$-modules with ${\Z}/2$ lowest weights}
Let $V_{H8}$ be a code VOA constructed from $H_8$ 
and $\{e^1,...,e^8\}$ the set of coordinate conformal vectors. 
In this subsection, we will study the fusion rules of 
irreducible $V_{H8}$-modules having lowest weights in $\hf{\Z}$. 
Let $W$ be an irreducible $V_{H8}$-module and $\be$ its $\st$-word. 
By Lemma 5.1, $\be\in H_8$. In particular, $|\be|\equiv 0 \pmod{4}$ 
and so the lowest weight of $W$ is in ${1\over 4}{\bf Z}$.   

If $L$ is an irreducible $V_{H8}$-module with integer lowest weight, 
then there is no $\st$ entry in a 
lowest weight row since the sum of entries in a lowest weight row 
is an integer and so we conclude $L=\sum_{\al} M_{\al}$ as 
$T$-modules.  Since the $\st$-word of $L$ is $(0^8)$, a $T$-module 
structure on $L$ determines the $V_{H8}$-module structure uniquely 
and so $L=M_{\al+H_8}$ for some $\al\in {\Z}_2^8$.  

Before we treat irreducible modules whose lowest weights are half-integers, 
we will prove the following lemma by the similar argument as in Theorem 6.12
in \cite{M1}.\\

\begin{lmm}  
Let $V=\sum_{i=0}^{\infty}V_i$ be 
a VOA over ${\R}$ satisfying $\dim V_0=1$ and $V_1=0$. 
Assume that $V$ has a positive definite invariant bilinear form 
$(\ast,\ast)$ and $e$ and $f$ are two   
distinct non-orthogonal rational conformal vectors with 
central charge $\half$.  Assume further that $\tau_e$ fixes $f$ 
and $\tau_f$ fixes $e$. 
If $L$ is a $V$-module such that $L$ is isomorphic to a direct sum 
of $L(\hf,\st)$ as $<\!e\!>$-modules and $<\!f\!>$-modules, 
then there is a conformal vector $d$ with central charge $\half$ such that
 $L$ has no $L(\hf,\st)$ as $<\!d\!>$-modules. 
\end{lmm}

\pr
As in the proof of lemma 6.11 in \cite{M1}, 
we have $\al,\be\in V_2$ satisfying 
$e\al=0$ and $e\be=\hf\be$ such that 
$ f=\la e+\al+\be. $
In the proof of Lemma 6.11 in \cite{M1}, we proved that $\la={1\over 8}$ 
and 
${16\over 5}\al$ is a conformal vector with central charge ${7\over 10}$. 
Since $\al_1$ and $e_1$ acts on $L$, so does $\be_1$. 
It is easy to see that $d={1\over 8}e+\al-\be$ is also a 
conformal vectors with central charge $\half$.
Since the lowest weights of ${16\over 5}\al$ is one of $0, {3\over 80}, 
{8\over 80}, {35\over 80}, {48\over 80}, {120\over 80}$ 
by the representation of $L({7\over 10},0)$, 
$d$ cannot have $\st$ as an eigenvalue on $L$.  Hence, 
$d$ is a conformal vector satisfying the desired conditions. 
\prend

If $L$ is an irreducible $V_{H8}$-module with half-integer lowest weight, 
then $L=(\sum_{\al}M_{\al})+(\sum \otimes_{i=1}^8 L^i(\hf,\st))$
as $T$-modules and so we have $L\cong M_{\al+H8}$ 
or $L\cong \sum \otimes_{i=1}^8L^i(\hf,\st)$ as $T$-modules by 
the fusion rules $(1.1)$. 
If $L\cong M_{\al+H_8}$ as $T$-modules, then $L\cong M_{\al+H_8}$ 
as $V_{H8}$-modules as in the case of integer lowest weight. 
If $L=\sum \otimes_{i=1}^8L^i(\hf,\st)$, then 
there is a set of mutually orthogonal 
conformal vectors, say $\{d^1,...,d^8\}$, such that 
$L$ is isomorphic to $M_{\al+H8}$ for some $\al\in {\Z}_2^8$ 
as $<\!d^1,...,d^8\!>$-modules by Lemma 6.4. 
So, we have the following classification. \\

\begin{thm}
If $L$ is an irreducible $V_{H8}$-module with half-integer or integer 
lowest \\
weight, then $L$ is isomorphic to one of the followings: \\
(1)  $M_{\nu_1+\nu_i+H8}$ for $i=1,...,8$ as $<\!e^1,...,e^8\!>$-modules, \\
(2)  $M_{\nu_i+H8}$ for $i=1,...,8$ as $<\!e^1,...,e^8\!>$-modules, \\
(3)  $M_{\nu_i+H8}$ for $i=1,...,8$ as $<\!f^1,...,f^8\!>$-modules, and \\
(4)  $M_{\nu_i+H8}$ for $i=1,...,8$ as $<\!d^1,...,d^8\!>$-modules. \\
Here $\nu_i$ denotes the word $(0,\cdots,0,1,0,\cdots,0)$ whose 
$i$th entry is $1$ and the others are all $0$.  
All irreducible modules in (3) and (4) are isomorphic to 
$\otimes_{i=1}^8 L(\hf,\st)$ as \\
$<\!e^1,...,e^8\!>$-modules.
\end{thm}

\pr
We have already proved that $L$ is isomorphic to one of the modules in the 
list.  We also showed that $M_{\nu_i+H_8}$ is isomorphic to 
$\otimes_{i=1}^8 L(\hf,\st)$ as $<f^1,...,f^8>$-modules in \cite{M2}. 
\prend

\begin{dfn}
We will use $H(h^i,\al)$ to denote the above modules, that is, \\
\begin{tabular}{ll}
 $H(\hf,\al)\cong M_{\al+H8}$ &as $<e^1,...,e^8>$-modules as in (1) and (2), \\
 $H(\st,\al)\cong M_{\nu_i+H8}$ &as $<f^1,...,f^8>$-modules as in (3) 
 if $f^i=s^{\al}$ and \\
 $H(\st,\be)\cong M_{\nu_i+H8}$ &as $<d^1,...,d^8>$-modules as in (4) 
 if $d^i=s^{\be}$, \\
\end{tabular} \\
where $s^{\al}$ is given in $(6.3)$.  Consequently, 
we have exactly thirty two irreducible $V_{H8}$-modules 
with lowest weights in ${\bf Z}/2$. 
\end{dfn}

The next lemma is clear by the definition.  

\begin{lmm} 
$H(h,\al)\cong H(k,\be)$ if and only if $h=k$ and $\al-\be\in H_8$. 
\end{lmm}

\begin{rmk}  The lowest weights of modules in $(2),(3),(4)$ are $\hf$ 
and those of the modules in $(1)$ are one. 
The modules $H(\st,\al)$ is characterized by the following properties: \\
(i)  $H(\st,\al)\cong \otimes L(\hf,\st)$ as $T$-modules. \\
(ii) The vertex operator of $u^{\be}$ on 
$H(\st,\al)$ is $(-1)^{\langle \al,\be\rangle}I(u^{\be},z)$. 
\end{rmk}

\subsection{Fusion rule of $V_{H8}$}
In this section, we will determine the fusion rules of the 
irreducible $V_{H8}$-modules with lowest weights in 
${\Z}/2$. In the previous section, we showed that 
such irreducible modules are 
$$ \left\{\ H(\hf,\al), \ H(\st,\be) \ :\ \al, \be\in {\Z}_2^8/H_8 
\right\}. \eqno{(6.6)}$$

We first prove the next lemma. \\

\begin{lmm}  If $\al$ is an even word, then 
$ H(\hf,\al)\cong M_{\al+H_8} $
as $<\!f^1,...,f^8\!>$-modules.  It is also true for $<\!d^1,...,d^8\!>$. 
\end{lmm}

\pr
By Lemma 6.3 and Definition 12, we have $\sigma_{e_1}(f^i)=d^i$ 
for all $i$ and $\sigma_{e_1}$ fixes all $H(\hf,\al)$. Hence, 
it is sufficient to prove the lemma for $<\!f^1,...,f^8\!>$. 
Without loss of generality, we may assume $\al=\{1,2\}$. 
The lowest weight of $H(\hf,\al)$ is one and the lowest weight 
space has a basis 
$\{ u^{\{12\}}, u^{\{34\}}, u^{\{56\}}, u^{\{78\}}\}$, where 
$\{1234\},\{1256\},\{1278\}$ are the set of four point codewords of 
$H_8$ containing $\{12\}$.  Hence, $H(\hf,\al)\cong M_{\be+H_8}$ 
as $<\!f^1,...,f^8\!>$-modules for some $\be$.  
By the direct calculation of the eigenvalues of 
$s^{\be}={1\over 8}{\bw}
+{1\over 8}\sum_{\ga\in H_8, |\ga|=4}(-1)^{<\be,\ga>}u^{\ga} $
on $H(\hf,\al)$, the eigenvalues of 
$s^{\nu_1}$ and $s^{\nu_2}$ on 
$-u^{\{12\}}+u^{\{34\}}+u^{\{56\}}+u^{\{78\}}$ are the same $\half$ and 
those of others are $0$.  Hence, $H(\hf,\{12\})=M_{\{12\}+H_8}$
as $<\!f^1,...,f^8\!>$-modules. 
\prend

By Theorem 5.5 and Theorem 6.2, we have the following theorem.

\begin{thm}  
If $W$ is an irreducible $V_{H8}$-module, then 
the tensor product $H(h,\be)\ts W$ is an irreducible module for any 
$h$ and $\be$. 
\end{thm}

\begin{rmk}
Let $v\in H(h,\be)$. 
By the direct calculation, we have 
$$ \begin{array}{l}
(z_1-z_2)^NY(s^{\al},z_1)Y(u^{\ga},z_2)v \\
={1\over 8}(z_1-z_2)^N
\left\{ Y(\hat{{\bf w}},z_1)Y(u^{\ga},z_2)+
\sum_{\be\in C, \ |\be|=4}(-1)^{(\al,\be)}Y(u^{\be},z_1)
Y(u^{\ga},z_2) \right\}v \\
={1\over 8}(z_1-z_2)^N\left\{ 
Y(u^{\ga},z_2)Y(\hat{{\bf w}},z_1)
+\sum_{\be\in C, \ |\be|=4} (-1)^{(\al+\ga,\be)}Y(u^{\ga},z_2)Y(u^{\be},z_1)
\right\}v \\
=(z_1-z_2)^NY(u^{\ga},z_2)Y(s^{\al+\ga},z_1)v 
\end{array} \eqno{(6.7)} $$
Hence, the action of $Y(s^{\al},z)$ on $Y(u^{\ga},z)H(h,\be)$ is same as 
$Y(s^{\al+\ga},z)$ on $H(h,\be)$.  So we have all actions of $s^{\al}$ on 
$H(\hf,\ga)\ts U$.  
\end{rmk}

As a corollary of Theorem 6.3 and Remark 8, we have:

\begin{cry}  The fusion rules of $V_{H8}$-modules 
$\{ H(\hf,\al), H(\st,\al) \}$
are \\
 $$ \begin{array}{ll}
  H(\hf,\al)\ts H(\hf,\be)=H(\hf,\al+\be), &\\
  H(\hf,\al)\ts H(\st,\be)=H(\st,\al+\be) &\mbox{ and }\\
  H(\st,\al)\ts H(\st,\be)=H(\hf,\al+\be).&
\end{array} \eqno{(6.8)} $$
\end{cry}

\pr
Changing the set of coordinate conformal vectors, it is sufficient 
to calculate the tensor product $H(\hf,\al)\ts H(h,\be)$. 
We have already proved that $H(\hf,\al)\ts H(h,\be)$ is 
an irreducible module by Theorem 5.5. 
Since the action of $Y(s^{\ga},z)$ on $Y(u^{\al},z)H(h,\be)$ is same as 
$Y(s^{\ga+\al},z)$ on $H(h,\be)$ by Remark 8, we have 
the desired fusion rules. 
\prend

\begin{thm} 
If $h=\hf$ and $\al$ is an even set and $\al\not\in H_8$, then 
$V_{H8}\oplus H(h,\al)$ has a simple VOA structure on it. 
If $h=\st$ or $\al$ is odd, then 
$V_{H8}\oplus H(h,\al)$ has a simple SVOA structure on it. 
\end{thm}

\pr  
Since the fusion rule $H(h,\al)\times H(h,\al)=V_{H8}$ has a 
single component, the structure is unique. 
If $h=\hf$ and $\al$ is even set, then 
$V=V_{H8}\oplus H(\hf,\al)$ is a subspace of $M_{S}$, where 
$S$ is the code consisting of all even words. 
Since $S$ is an even linear code, $M_{S}$ has a VOA structure on it with 
positive definite invariant bilinear form, so does $V$. 
If $h=\st$ or $\al$ is odd, then we may assume that 
$h=\hf$ and $\al$ is odd by Definition 13. For an odd word $\al$,  
$V_{H8}\oplus H(\hf,\al)$ is a subspace of $M_{{\Z}_2^n}$ and 
$M_{S}\cap (V_{H8}\oplus H(\hf,\al))=V_{H8}$.  
Since $M_{{\Z}_2^n}$ has a SVOA structure on it with the even part 
$M_{S}$ and the odd part $M_{S+\al}$, we have the desired conclusions.  
\prend
 
\subsection{VOA structure and fusion rules}
 
In this subsection, $E$ denotes a self-orthogonal subcode of $D$ and 
we assume that $E$ is a direct sum $E=\oplus E^i$ of Hamming codes $E^i$ 
and $U=\otimes H(h^i,\al^i)$ is an $M_E$-module such 
that $M_E\oplus U$ has a simple VOA structure.  
We next show that the induced $M_D$-module also has a simple VOA structure.  

\begin{thm}  If $M_E\oplus U$ has a simple VOA structure on it, then 
so does $M_D\oplus Ind_{M_E}^{M_D}(U)$.
\end{thm}

\pr Let $(M_D,Y^0)$ be a code VOA and $(M_E\oplus U, Y^1)$ a simple VOA.  
Since $U\times U=M_E$ and $U\times M_E=U$, 
the vertex operator $Y^1(u,z)$ of $u\in U$ has an 
expression 
$$Y^1(u,z)=\pmatrix{0 &I^1(u,z) \cr I^2(u,z) & 0},\eqno{(6.9)}$$
where 
$I^1(u,z)\in {\rm Hom}(U,M_E)\{z,z^{-1}\}$, and 
$I^2(u,z)\in {\rm Hom}(M_E,U)\{z,z^{-1}\}$. \\
Set $W=Ind_{M_E}^{M_D}(U)$ and let $Y^W$ be an induced 
module vertex operator.  We first show $W\times W=M_D$.  
Let $U^0$ be an $M_D$-module with $I_{M_D}\pmatrix{U^0\cr W\quad W}\not=0$.
Then $I_{M_E}\pmatrix{U^0\cr U\quad U}\not=0$ by Theorem 5.4 
and so $U^0$ contains 
$M_E=U\times U$ as $M_E$-modules. Since the $M_D$-module structure 
of $U^0$ is determined by an $M_E$-submodule by Theorem 5.3, 
we see $U^0=M_D$.  
Since $M_E$-module $M_D$ has only one irreducible submodule 
isomorphic to $M_E$, 
we have $W\times W=M_D$ and 
there is a non-zero intertwining operator 
$I(\ast,z)\in I\pmatrix{M_D\cr W\quad W}$. 
We will next define vertex operators 
$$Y(v,z)\in {\rm End}(M_D\oplus Ind_{M_E}^{M_D}(U))[[z,z^{-1}]]$$
of $v\in W=Ind_{M_E}^{M_D}(U)$  
satisfying the derivation and 
the mutually commutativity. 
We also have an intertwining operator $I'(\ast,z)\in 
I\pmatrix{W\cr W\quad M_D}$ since $W\times M_D\cong M_D\times W$.  
We therefore obtain a vertex operator 
$$Y(v,z)=\pmatrix{0 & I(v,z) \cr I'(v,z) &0}\in 
{\rm End}(M_D\oplus W)[[z,z^{-1}]]
\eqno{(6.10)}$$ 
of $v\in W$.  By the properties of intertwining operators, 
$Y(v,z)$ satisfies the derivation and 
the commutativity and the associativity with the vertex operators 
$$ Y(u,z)=\pmatrix{Y^0(u,z) &0 \cr 0&Y^W(u,z)} \eqno{(6.11)}$$
of $u\in M_D$.  By the construction, $M_E$-module ${\rm Ind}_{M_E}^{M_D}(U)$ 
contains only one irreducible submodule isomorphic to $U$ and $U\times U=M_E$. 
Hence, if we restrict this vertex operator to that of $v\in M_E\oplus U$, then 
it should be equal to the vertex operator 
$Y^1(v,z)\in {\rm End}(M_E\oplus U)[[z,z^{-1}]]$ of $v\in M_E\oplus U$ 
by multiplying some scalars to $I(v,z)$ and $I'(v,z)$. In particular, 
there is a sufficiently large integer $N$ such that 
$$(z_1-z_2)^NY(v,z_1)Y(v,z_2)w=(z_1-z_2)^NY(v,z_2)Y(v,z_1)w \eqno{(6.12)}$$  
for $w\in M_E\oplus U$. 
Since $Y(v,z)$ satisfies the commutativity with $Y(u,z)$  
of elements $u\in M_D$, we have 
$$ \begin{array}{l}
(z_1-z_2)^N(z_1-z_3)^N(z_2-z_3)^NY(v,z_1)Y(v,z_2)Y(u,z_3)w \cr
=(z_1-z_2)^N(z_1-z_3)^N(z_2-z_3)^NY(u,z_3)Y(v,z_1)Y(v,z_2)w \cr
=(z_1-z_2)^N(z_1-z_3)^N(z_2-z_3)^NY(u,z_3)Y(v,z_2)Y(v,z_1)w \cr
=(z_1-z_2)^N(z_1-z_3)^N(z_2-z_3)^NY(v,z_2)Y(v,z_1)Y(u,z_3)w 
\end{array}\eqno{(6.13)}$$
for a sufficiently large integer $N$. 
Hence, by multiplying a power of $z_3$ and substituting $0$ into $z_3$, 
we obtain 
$$ (z_1-z_2)^NY(v,z_1)Y(v,z_2)u_rw=(z_1-z_2)^NY(v,z_2)Y(v,z_1)u_rw,
\eqno{(6.14)}$$  
where $u_rw\not=0$ and $u_{r+i}w=0$ for all $i>0$.  
Substituting this into the above, we see  
$$ (z_1-z_2)^NY(v,z_1)Y(v,z_2)u_iw=(z_1-z_2)^NY(v,z_2)Y(v,z_1)u_iw 
\eqno{(6.15)}$$  
for all $i$. Since $\{ u_iw: u\in M_D, w\in M_E\oplus U\}$ 
spans whole space $M_D\oplus W$ by Proposition 4.1 in \cite{DM}, 
$Y(v,z)$ satisfies the commutativity with itself. 
It is clear that it satisfies the other conditions required to be 
a vertex operator of VOA.  We note that we don't need to prove 
the associativity of $Y(\ast,z)$. 
It is also clear that $(V,Y)$ is simple. 
\prend

Let $X$ be an irreducible $M_D$-module and assume that an irreducible 
$M_E$-submodule of $X$ is isomorphic to 
$\otimes_{i=1}^k H(h^i,\be^i)$ for some 
$h^i\in \{\hf, \st\}$ and $\be^i\in {\bf Z}_2^8$, where $n=8k$. 
We will calculate the fusion rules of $X$ with others.  Let $\al^1$ be 
an $\st$-word of $X$.  In order to simply the notation, we assume 
$\al^1=(1^{8s}0^{8(k-s)})$.  Namely, we will assume the following:  \\

\noindent
{\bf Hypotheses I} \\
(1)\quad Let $D$ be an even binary linear code with length $8k$ 
and $E$ a self-orthogonal subcode of $D$. \\
(2)\quad Assume that $E$ is a direct sum 
$\oplus_{i=1}^k E^i$ of Hamming codes $E^i$. \\
(3)\quad Let $U$ be an $M_E$-module 
$(H(\st,\al^1)\otimes \cdots \otimes H(\st,\al^s))\otimes 
(H(\hf,\al^{s+1}\otimes \cdots \otimes H(\hf,\al^k)). $ \\
\mbox{}\qquad and set $M^1={\rm Ind}_{M_E}^{M_D}(U)$.  We note that 
$\al^1=(1^{8s}0^{8(k-s)})$ is a $\st$-word of $M^1$. \\
(4)\quad Let $M^2$ be an irreducible $M_D$-module with a $\st$-word $\al^2$. 
Set $\al^3=\al^1+\al^2$.  \\ 
(5) \quad Set $K^i=\{\be\in D| \be\subseteq \al^i\}$ and assume 
$E+K^2=E+K^3$.  We note that 
$K^1$ contains $\oplus_{i=1}^sE^i$.   \\

\begin{rmk} We will explain the meaning of the last assumption.
Choose $\al\in K^2$, then there is an $\be_{\al}\in E$ such that 
$\al+\be_{\al}\in K^3$.  Set $\ga=\al^1\cap \al^2$, 
$\ga^1=\al^1-\ga$ and $\ga^2=\al^2-\ga$.  
The above relation implies that 
$\al\cap \tilde{h}=\be_{\al}\cap \tilde{h}$ since 
$(\al+\be_{\al})\cap \ga=\emptyset$.  
Since $\al+\be_{\al}\in K^3$ and $\al^1+\la^2\cap \ga=\emptyset$, 
we have $\al+\be_{\al}\subseteq \al^3=\al^1+\al^2$ and so 
$\al+\be_{\al}\cap \ga=\emptyset$. 
If $\al\in K^2$, then $\al=(\al\cap\ga)\cup (\al\cap\ga^2)$
and $\ga\cap \ga^1=\emptyset$.  Hence, 
for $\al,\xi\in K^2$,  we obtain 
$$<\al,\ga>=\langle \al\cap \ga,\xi\cap \ga\rangle
+\langle \al\cap \ga^2,\xi\cap \ga^2\rangle. $$
Similarly, we have 
$$ 
0=<\be_{\al},\be_{\xi}>
=\langle \be_{\al}\cap \ga^1,\be_{\xi}\cap \ga^1\rangle
+\langle \be_{\al}\cap \ga,\be_{\ga}\cap \ga\rangle 
$$
for $\be_{\al},\be_{\xi}\in E$ since $E$ is self-orthogonal. 
Hence, if 
$\langle \al,\ga\rangle=0$ for $\al,\xi\in K^2$, then 
$$\langle \al\cap \ga,\xi\cap \ga\rangle
=\langle \al\cap \ga^2,\xi\cap \ga^2\rangle $$
and 
$$ \langle \be_{\al}\cap \ga^1,\be_{\xi}\cap \ga^1\rangle
=\langle \be_{\al}\cap \ga,\be_{\ga}\cap \ga\rangle. $$
Therefore, we obtain
$$ \begin{array}{ll}
\langle \al+\be_{\al},\xi+\be_{\xi}\rangle 
&=\langle \al\cap \ga^2+\be_{\al}\cap \ga^1,
\xi\cap \ga^2+\be_{\xi}\cap \ga^1\rangle \\
&=\langle \al\cap \ga^2,\xi\cap \ga^2\rangle+
\langle \be_{\al}\cap \ga^1,\be_{\xi}\cap \ga^1\rangle=0. \\
\end{array} $$
Namely, if $H^2$ is a self-orthogonal subcode of $K^2$, then 
$\{ \be_{\al}:\al\in H^2 \}$ is also a self-orthogonal.  
In particular, if $H^2$ is a maximal self-orthogonal subcode of $K^2$, 
then there is a maximal self-orthogonal subcode $H^3$ of $K^3$ such that 
$E+H^2=E+H^3$.  
\end{rmk}

We will prove the one of our main theorems. 

\begin{thm} Under Hypotheses I, $M^1\times M^2$ is irreducible.  
\end{thm}

\pr 
Let $M^3$ be an irreducible $M_D$-module such that 
$I_{M_D}\pmatrix{M^3\cr M^1\quad M^2}\not=0$ and   
$U^2$ an irreducible $M_E$-submodule of $M^2$. 
Clearly, $\al^3$ is the $\st$-word of $M^3$.
Let $H^i$ 
be maximal self-orthogonal subcode of $K^i$ such that $E+H^2=E+H^3$ 
by Remark 9.   
Assume first that $D=E+H^3$.  Then 
$M^2$ and $M^3$ are both irreducible $M_E$-modules by Theorem 5.1 
and we have 
$$\dim I_{M_D}\pmatrix{M^3\cr M^1\quad M^2}
\leq \dim I_{M_E}\pmatrix{M^3\cr U \quad M^2}=1 $$
by Theorem 5.4. 
We therefore obtain $M^3=U\times M^2$ as $M_E$-modules.
Choose a nonzero intertwining operator 
$I^1(\ast,z)\in I_{M_E}\pmatrix{M^3\cr U\quad M^2}$. 
Then for $I(v,z)\in I_{M_D}\pmatrix{M^3\cr M^1\quad M^2}$, there is 
a scalar $\la$ such that 
$I(v,z)=\la I^1(v,z)$ for $v\in U$. 
By the commutativity of intertwining operator 
$$0=(z_1-z_2)^NY^3(u,z_1)I(v,z_2)w=(z_1-z_2)^NI(v,z_2)Y^2(u,z_1)w,  $$
we have 
$$0=(z_1-z_2)^N\la Y^3(u,z_1)I^1(v,z_2)w=(z_1-z_2)^N\la I^1(v,z_2)Y^2(u,z_1)w$$
for all $v\in U$, $u\in M_D$ and $w\in M^2$ and for a sufficiently large 
integer $N$.  
Hence, the action $Y^3(u,z)$ of $u\in M_D$ on $M^3$ 
does not depend on the choice 
of $\la$ and it is determined by $Y^2(u,z)$ and $I^1(v,z_2)$.  
Therefore, the structure of $M^3$ is uniquely determined and 
$M^1\times M^2$ is irreducible.  

We will next prove the general case. Assume $H=E+H^3\not= D$.  
Let $U^1$ and $U^2$ be irreducible $M_H$-submodules of 
$M^1$ and $M^2$, respectively.  Then the tensor product 
$U^1\times U^2$ is an irreducible 
$M_H$-module as we showed and hence $M^3$ contains $U^1\times U^2$ 
as $M_H$-submodule. 
Therefore the $M_D$-module structure on $M^3$ is uniquely determined 
by Theorem 5.3 and so $M^1\times M^2$ 
is irreducible. 
\prend

\section{Construction and uniqueness of VOA structure}
In this section, we will show a new construction of 
vertex operator algebras.  We will define a Fock space $V$ 
containing $M_D$ and then show that the vertex operators of elements 
will be automatically determined by its representations.

\subsection{The setting}
We will assume the following Hypotheses II $(1)\sim (8)$.  \\

\noindent
{\bf Hypotheses II }\\
(1) \quad $D$ and $S$ are both even linear codes with length $8k$. \\
(2) \quad $\langle D, S\rangle=0 $. \\
(3) \quad For all $\al\in S$, the weight $|\al|$ is a 
multiple of eight. \\
(4) \quad For any $\al\in S$, $D$ contains a self-dual subcode $E_{\al}$ 
such that $E_{\al}$ is a direct sum 
$E_{\al}=\oplus_{i=1}^kE^i_{\al}$ 
of Hamming codes $E_{\al}^i$.  Assume that 
$H_{\al}=\{\be\in E_{\al}:\be\subseteq \al\}$ is a 
direct factor of 
$E_{\al}$ containing $\al$. \\
(5) \quad Set $K_{\al}=\{\be\in D:\be\subseteq \al\}$ and assume that 
for any two $\al,\be\in S$, $K_{\al}+H_{\be}=K_{\al+\be}+H_{\be}$. \\ 

\noindent
Let $\{\al^1,...,\al^t\}$ be a basis of $S$.  For each $j$, 
$E_{\al^j}$ is a direct sum $\oplus_{i=1}^k E^i_{\al^j}$ of Hamming codes 
and 
$U_{\al^j}=\otimes_{i=1}^k H(h^j_i,\be^j_i)$ is an $M_{E_{\al^j}}$-module 
with a $\st$-word $\al^j$. \\
$(6)$ \quad We assume that $M_{E_{\al^j}}\oplus U_{\al^j}$ has a simple VOA 
structure on it. \\

As we showed in Theorem 6.4, if $\be$ is even, then 
$V_{H8}\oplus H(\hf,\be)$ has a VOA structure and 
if $h=\st$ or $\be$ is odd, then $V_{H8}\oplus H(h,\be)$
 has a SOVA structure. 
Hence, $(6)$ is equivalent to :\\
$(6')$ \quad $|\{i :h^j_i=\st, \al^j_i \mbox{ is odd } \}|\equiv 0 \pmod{2}$ 
for all $j$. \\

Set 
$$ V^{\al^i}=Ind_{M_{E^i}}^{M_D}(U_{\al^i}).  $$
This is an irreducible $M_D$-module.  By Theorem 6.5, 
$M_D\oplus V^{\al^i}$ has a simple VOA structure. \\

By Theorem 6.6 and the above hypotheses, 
$V^{\al^i}\times V^{\al^j}$ is irreducible. 
Set $V^{\al^i+\al^j}=V^{\al^i}\times V^{\al^j}$.  
By the symmetry of the fusion rule and the 
uniqueness, it is equal to $V^{\al^j+\al^i}$. In order to simplify 
the notation, we will denote $\al^i+\al^j$ by $\al$ for a while. 
The $\st$-word of 
$V^{\al}$ is $\al$ and 
there is a self-orthogonal subcode $E_{\al}$ in $D$ such that $E_{\al}$ 
is a direct sum of Hamming codes  $E^i_{\al}$ by 
the condition (4). Hence $V^{\al}$ is 
also an induced module $Ind_{M_{E_{\al}}}^{M_D}(U_{\al})$ for 
an irreducible $M_{E_{\al}}$-module $U_{\al}\cong \otimes H(h^j,\be^j)$.  
We can therefore define the tensor product 
$V^{\al^k+\al^i+\al^j}=V^{\al^i+\al^j}\times V^{\al^k}$ uniquely. 
Repeating these steps, 
we can define all $M_D$-modules $V^{\al}$ for $\al\in S$ by 
$$  V^{\al}=(\cdots(V^{\al^{j_1}}\times V^{\al^{j_2}})\times \cdots ) 
\eqno{(7.1)} $$
for $\al=\al^{j_1}+\cdots +\al^{j_r}$ with $j_1<\cdots <j_r$.  \\

\noindent
We assume the associativity: \\
(7) \quad $V^{\al}$ does not depend on the order of products.  \\

It is easy to see that $(7)$ is equivalent to 
the associativity of product of three elements: 
$$  (V^{\al^i}\times V^{\al^j})\times V^{\al^k}
=V^{\al^i}\times (V^{\al^j}\times V^{\al^k}).  \eqno{(7.2)} $$

By the definition of $V^{\al}$ and the assumption (7), 
we have:

\begin{lmm} 
$V^{\al}\times V^{\be}=V^{\al+\be}$.
\end{lmm}

\vspace{5mm}

At last, we assume that the commutativity of intertwining operators: \\

\noindent
(8) \quad For $I^{\be}(\ast,z)
\in I\pmatrix{V^{\al+\be}\cr V^{\al}\quad V^{\be}}$ 
and $I^{\be+\al}(\ast,z)\in I\pmatrix{V^{\be}\cr V^{\al}\quad V^{\al+\be}}$,  
we assume that the powers of $z$ 
in $I^{\be}(\ast,z)$ and $I^{\be+\al}(\ast,z)$ 
are all integers and they satisfy the commutativity, that is, 
for $v\in V^{\al}$, we assume  
$$  (z_1-z_2)^NI^{\be}(v,z_1)I^{\be+\al}(v,z_2)
=(z_1-z_2)^NI^{\be+\al}(v,z_2)I^{\be}(v,z_1)  \eqno{(7.3)}$$
for sufficiently large integer $N$.  \\

Set 
$$  V=\oplus_{\al\in S}V^{\al}, \eqno{(7.4)}$$ 
where $V^{0}=M_D$.

Under the above Hypotheses II $(1)\sim (8)$, we will show that 
a vertex operator $Y(v,z)$ for every element $v\in V$ is automatically 
determined.

\subsection{Construction of vertex operators}
Let $\dim S=t$ and 
set $S_i=<\!\al^1,...,\al^i\!>$ for $i=0,1,...,t$. 
Set $V^{i}=\oplus_{\al\in S_i}V^{\al}$. 
We will define a vertex operator $Y(v,z)\in {\rm End}(V)[[z,z^{-1}]]$ of 
$v\in V^i$ inductively.  Since $V^{\al}$ are all $M_D$-modules, 
the vertex operators $Y(v,z)$ of $v\in V^0=M_D$ on $V$ are already 
determined and 
they satisfy the mutual commutativity. 
Assume the vertex operators $Y(w,z)\in {\rm End}(V)[[z,z^{-1}]]$ 
of elements $w\in V^r$ 
are already determined and they satisfy the mutual commutativity and 
we will define $Y(v,z)\in {\rm End}(V)[[z,z^{-1}]]$ for $v\in V^{r+1}$.   
Decompose $V^{r+1}=V^r\ops W$ as $V^r$-modules, then 
$W$ is an irreducible $V^r$-modules and 
$W=\oplus_{\be\in S_{r+1}-S_r}V^{\be}$.  
Set $\al=\al^{r+1}\in S_{r+1}-S_r$ to simplify the notation. 
By the assumption $(6)$ and Theorem 6.5, 
$M_D\oplus V^{\al}$ has a simple VOA structure 
and denote it by 
$$ (M_D\oplus V^{\al}, Y^{\al}). \eqno{(7.5)} $$

We first prove:

\begin{lmm}  $W\times W=V^r$ as $V^r$-modules. 
\end{lmm}

\pr 
Let $U$ be an irreducible $V^r$-submodule of $W\times W$. By the same 
arguments as in the proof of Theorem 5.4, we have  
$$ \dim I_{V^r}\pmatrix{U\cr W\quad W} \leq 
\dim I_{M_D}\pmatrix{U\cr V^{\al} \quad V^{\al+\be}}  \eqno{(7.6)}$$  
for any $\be\in S_r$ 
and hence $U$ contains $V^{\al}\times V^{\al+\be}=V^{\be}$ as 
$M_D$-modules. Therefore the $V^r$-structure on $U$ is uniquely determined 
and so $U\cong V^r$.  
\prend

By the same arguments as in the above proof, 
$\oplus_{\de\in S_r}V^{\de+\mu}$ is an irreducible $V^r$-module and 
we have: 
$$ W\times (\oplus_{\de\in S_r}V^{\de+\ga})
=\oplus_{\de\in S_r}V^{\de+\ga+\al}, \eqno{(7.7)}$$
which is also an irreducible $V^r$-module.  
As we showed in the proof of Theorem 5.4, we can induce 
every intertwining 
operator in  \\
$I_{V^r}\pmatrix{\oplus_{\de\in S_r}V^{\de+\al+\ga} \cr
W\quad \oplus_{\de\in S_r}V^{\de+\ga}}$ 
from one in $I_{M_D}\pmatrix{V^{\de+\al}\cr V^{\al}\quad V^{\de}}$.
Namely, we can choose a nontrivial intertwining operator \\
$I^{\ga}(\ast,z)\in I_{V^r}\pmatrix{\oplus_{\de\in S_r}V^{\de+\al+\ga}\cr
W\quad \oplus_{\de\in S_r}V^{\de+\ga}}$ for each 
$\ga\in S/S_r$ such that \\
$I^{\ga}(\ast,z)_{|V^{\ga}}\in I_{M_D}\pmatrix{V^{\al+\ga} \cr
V^{\al}\quad V^{\ga}}$ and $I^{\ga}(\ast,z)_{|V^{\ga+\al}}
\in I_{M_D}\pmatrix{V^{\ga}\cr V^{\al}\quad V^{\al+\ga}}$. \\
Since $I^{\ga}(\ast,z)$ is an intertwining operator, it satisfies 
the commutativities: \\
$$  (z_1-z_2)^NY^{\ga+\al}(u,z_1)I^{\ga}(v^{\al},z_2)w=
(z_1-z_2)^NI^{\ga}(v^{\al},z_2)Y^{\ga}(u,z_1)w  \eqno{(7.8)} $$
for $v^{\al}\in V^{\al}$, $u\in V^r$, 
$w\in \oplus_{\delta\in S_r} V^{\ga+\delta}$ and 
$$  (z_1-z_2)^NY^{\ga}(u,z_1)I^{\ga+\al}(v^{\al},z_2)w'=
(z_1-z_2)^NI^{\ga+\al}(v^{\al},z_2)Y^{\ga+\al}(u,z_1)w'   \eqno{(7.9)}$$
for $v^{\al}\in V^{\al}$, $u\in V^r$, 
$w'\in \oplus_{\delta\in S_r} V^{\ga+\al+\delta}$. 
Since a VOA $(M_D\oplus V^{\al},Y^{\al})$ is given, 
$I^{\al}(\ast,z)\in I\pmatrix{W\cr W\qquad V^r}$ 
is uniquely determined by the property:
$$v^{\al}_{-1}{\bf 1}=v^{\al},$$ where 
$I^{\al}(v^{\al},z)=\sum v^{\al}_nz^{-n-1}$ for $v^{\al}\in V^{\al}$.   
Also, by the commutativity of $Y^{\al}$ on $M_D\oplus V^{\al}$, 
$I^0(v^{\al},z)\in I\pmatrix{V^r \cr W\quad W}$  
is uniquely determined.

Define a vertex operator 
$Y(v^{\al},z)\in {\rm End}(V)[[z,z^{-1}]]$ of $v^{\al}\in V^{\al}$  
by $$ Y(v^{\al},z)=\left(
\begin{array}{llll}
\begin{array}{ll}
0 &I^0(v^{\al},z) \cr
I^{\al}(v^{\al},z) & 0
\end{array}  &  &  &  \cr
&\begin{array}{ll}
0 &I^{\ga}(v^{\al},z) \cr
I^{\ga+\al}(v^{\al},z) & 0
\end{array} & & \cr
& & \cdots & \cr
& & & \cdots \cr
\end{array} \right),  \eqno{(7.10)}$$
then it satisfies the commutativity with 
$Y(w,z)$ for all $w\in V^r$ as we showed. 
Moreover, the assumption $(8)$ means the commutativity: 
$$ (z_1-z_2)^NY(v^{\al},z_1)Y(v^{\al},z_2)w
=(z_1-z_2)^NY(v^{\al},z_2)Y(v^{\al},z_1)w \eqno{(7.11)}$$
for any $\al\in S, w\in V^{\be}$.
Therefore, $Y(v^{\al},z)$ satisfies the commutativity with itself. 
Since every $I^{\ga}(v^{\al},z)$ satisfies the 
derivation, so does $Y(v^{\al},z)$.
Using the normal product, we can define all vertex operators $Y(v,z)$ 
of $v\in V^{r+1}$, which still satisfy the mutual commutativity and 
the derivation. We therefore have a VOA structure on $V$.  
We next show that the VOA structures on $V$ is unique.   
Clearly, if $V$ has a VOA structure containing $M_D$ as a subVOA, 
then $V$ has the above structure that we have constructed. Since 
we can modify the difference of scalar times of 
$I^{\ga}(v^{\al},z)$ by multiplying the scalar to the bases of 
$\oplus_{\de\in S_{r}}V^{\de+\ga}$ and 
$\oplus_{\de\in S_{r}}V^{\de+\ga+\al}$, we obtain the uniqueness of 
vertex operators up to the change of basis.  

This completes the construction of VOA.

\begin{thm}  Under the hypotheses II $(1)\sim (8)$, 
$V$ has a unique VOA structure.
\end{thm}

\begin{rmk}  Except the last two assumptions, they are conditions on 
the codes.  So if we are given a pair of codes $D$ and $S$ 
satisfying the conditions $(1)\sim (5)$ and choose the desired 
$M_D$-modules $V^{\al}$, what we have to do is to check the 
associativity $(7)$ and the commutativity $(8)$.  
We should note that these conditions are possible to make sure 
among modules generated by three modules $V^{\al}, V^{\be}, V^{\ga}$.  
Sometimes, these information are possible to 
get from the known VOA.  Namely, if 
there is a VOA $V$ containing a set of mutually orthogonal 
rational conformal 
vectors $\{e^i:i=1,...,n\}$ with central charge $\hf$ 
whose sum is the Virasoro element, then we obain 
the associativity and the commutativity in the decomposition 
$V=\oplus_{\chi\in Irr(P)} V_{\chi}.$ 
\end{rmk}

\end{document}